\documentclass[11pt]{article}
\usepackage{amsmath,amsthm,amssymb,graphicx}

\usepackage[width=1.00\textwidth,indention=0mm,justification=centerlast,
font=small,labelfont=bf,labelsep=period]{caption}

\usepackage[pdfpagemode=UseOutlines,pdfstartview={XYZ 1 0 1},pdfpagelayout=SinglePage,
bookmarksnumbered=true,bookmarksopen=true,bookmarksopenlevel=2,unicode=true,
pdfhighlight=/P,colorlinks=true,linkcolor=blue,urlcolor=blue,citecolor=blue]
{hyperref}

\advance\textwidth 20mm  \advance\oddsidemargin -10mm
\advance\textheight 30mm \advance\topmargin -18mm

\allowdisplaybreaks[1]

\theoremstyle{plain}
\newtheorem{theorem}{Theorem}
\newtheorem{statement}[theorem]{Statement}
\newtheorem{corollary}[theorem]{Corollary}
\theoremstyle{remark}
\newtheorem{remark}{Remark}
\theoremstyle{definition}
\newtheorem{definition}{Definition}

\newcommand{\gr}[1]{\centerline{\includegraphics[scale=0.45]{#1.pdf}}}
\newlength{\slen}
\newcommand{\sout}[1]{\settowidth{\slen}{$#1$}%
\hbox to 0pt{\rule[0.5ex]{\slen}{0.5pt}\hss}#1}
\newcommand{\supp}{\mathop{\rm supp}\nolimits}
\newcommand{\Real}{\mathbb R}
\newcommand{\StirlingII}[2]{\genfrac{\{}{\}}{0pt}{1}{#1}{#2}}
\newcommand{\Euler}[2]{\genfrac{\langle}{\rangle}{0pt}{1}{#1}{#2}}
\newcommand{\pos}[1]{\langle#1\rangle}
\newcommand{\hrefo}[1]{\href{http://oeis.org/#1}{#1}}
\newcommand{\citeo}[1]{\cite[\hrefo{#1}]{OEIS}}

\title{Set partitions and integrable hierarchies}
\author{V.E. Adler\footnote{L.D. Landau Institute for Theoretical Physics,
Ac.~Semenov str.~1-A, 142432 Chernogolovka, Russian Federation. E-mail:
adler@itp.ac.ru}}

\date{2 October 2015}

\begin{document}\maketitle

\begin{abstract}
We demonstrate that statistics for several types of set partitions are
described by generating functions which appear in the theory of integrable
equations.
\smallskip

\smallskip

\end{abstract}

\section{Introduction}\label{s:intro}

In combinatorics, a typical problem is to establish algorithms for generating
and counting of all objects of a given type. We are interested in the situation
when:

--- such an algorithm can be described in terms of generating operations,
which make possible to build the objects under scrutiny recursively from the
objects with lesser number of elements;

--- each object can be associated with a number or a monomial in formal
variables, in such a way that the generating operations correspond to certain
algebraic or differential operations.

As a result, the set of all objects is mapped into a sequence of numbers or
polynomials governed by certain recurrence relations. A remarkable fact is that
in some cases the generating functions for these sequences appear also in the
theory of integrable equations. We mention the relation between the number
partitions and statistical mechanics \cite{Vershik_1996}, the combinatorics of
Painlev\'e transcendents \cite{Deift_2000}, of KdV solitons
\cite{Grosset_Veselov_2005} and of KP tau-function
\cite{Alexandrov_Mironov_Morozov_Natanzon_2012}, as just few examples of
interconnection of both theories. A well known example, which is more close to
the subject of this paper, is given by the Bell polynomials which describe the
statistics of set partitions and also define the potential Burgers hierarchy
\cite{Lambert_Loris_Springael_Willox_1994, Lambert_Springael_2008}. Our goal is
to generalize this observation for several special types of set partitions with
restrictions on the structure of blocks.

In sections \ref{s:SP}, \ref{s:B} we consider the generic set partitions and
$B$ type partitions related, respectively, with the Burgers and
Ibragimov--Shabat linearizable hierarchies. Section \ref{s:n-seq} is the only
one, where set partitions are replaced by other combinatorial objects; the
corresponding equations are related with the Burgers hierarchy by the hodograph
transformation. In all these cases the definitions of monomials rely upon
counting of elements in blocks.

In section \ref{s:NOP}, we study the non-overlapping partitions associated with
the KdV hierarchy; section \ref{s:AP} is devoted to the atomic partitions and
the Kaup--Broer hierarchy (a gauge equivalent version of the nonlinear
Schr\"odinger hierarchy). Here, the combinatorics becomes more complicated, as
well as its statistics: the definition of the monomial corresponding to a
partition is based on the sequence of operations which bring to this partition,
rather than on the size of its blocks. Moreover, the combinatorics becomes more
disguised: in the Burgers case, the generating function is related
intermediately with the higher flows, while in the KdV and NLS cases it becomes
the asymptotic series in powers of the spectral parameter for the logarithmic
derivative of $\psi$-function, governed by Riccati equation.

The structure of all sections is uniform: first, we define the generating
operations for the combinatorial objects under consideration, next, we
introduce the polynomials which describe their statistics and derive the
recurrence relations, finally, a comparison with an integrable hierarchy is
given.

\section{Set partitions}\label{s:SP}

\subsection{Basic notions}\label{s:def}

Recall that a {\em set partition} is a set of nonempty, pairwise disjoint
subsets ({\em blocks}) of the set, such that their union gives the whole set.
The notation $\pi\vdash[n]$ means that $\pi$ is a partition of the set
$[n]=\{1,\dots,n\}$. The set of all partitions of $[n]$ is denoted $\Pi_n$ and
its subset consisting of all partitions with $k$ blocks is denoted $\Pi_{n,k}$,
$1\le k\le n$. As a rule, we denote the blocks by the same letter as the
partition, with a subscript: $\pi=\{\pi_1,\dots,\pi_k\}$. For $n=0$, it is
convenient to accept that the set $[0]=\varnothing$ admits the unique partition
$\varnothing$ which contains no blocks, that is,
$\Pi_0=\Pi_{0,0}=\{\varnothing\}$.

Let $|A|$ denote the number of elements of the set $A$. Then $|\pi|$ is the
number of blocks in the set partition $\pi$, and $|\pi_j|$ is the size of the
block $\pi_j$. A block is called a singleton if $|\pi_j|=1$, and a multiplet if
$|\pi_j|>1$. The total number of elements in the partitioned set will be
denoted $\|\pi\|$.

By definition, a partition is an unsorted set, that is, the order of blocks and
order of elements in blocks do not matter. Nonetheless, the ordering defined by
the enumeration of elements can be used for a definition of unique
representation of partitions. In the {\em canonical form}, a block of a
partition is represented by a list of elements sorted by increase, and the
partition itself is represented by a list of blocks, sorted by increase of
their first (minimal) elements. For short, it is common to write, for instance,
$\pi=1,2,7|3,5|4,9,12|6|8,10,11$, with the blocks separated by vertical bars.

Alternatively, a set partition in the canonical form is represented by the
sequence $s=(s_1,\dots,s_n)$, where $s_m$ is the index of the block containing
the element $m$; in the above example, it is $(1,1,2,3,2,4,1,5,3,5,5,3)$. This
gives rise to the integer sequences characterized by the {\em restricted
growth} condition:
\begin{equation}\label{rgs}
 s_1=1,\quad 1\le s_m\le\max(s_1,\dots,s_{m-1})+1.
\end{equation}
The set of such sequences of length $n$ will be denoted $R_n$.

Throughout the paper, we use the graphical representation of set partitions as
illustrated by fig.~\ref{fig:P-djM}. The elements are enumerated from left to
right, the rows represent the blocks sorted from top to bottom by increase of
minimal elements. The line between the outermost elements of a block is called
its {\em support}: $\supp\pi_j=[\min\pi_j,\max\pi_j]_\Real$. This notion is
used in sections \ref{s:NOP}, \ref{s:AP}, devoted to the so-called {\em
non-overlapping} and {\em atomic} partitions.

\subsection{Generating operations and generating functions}\label{s:gf}

\begin{figure}[t]
\gr{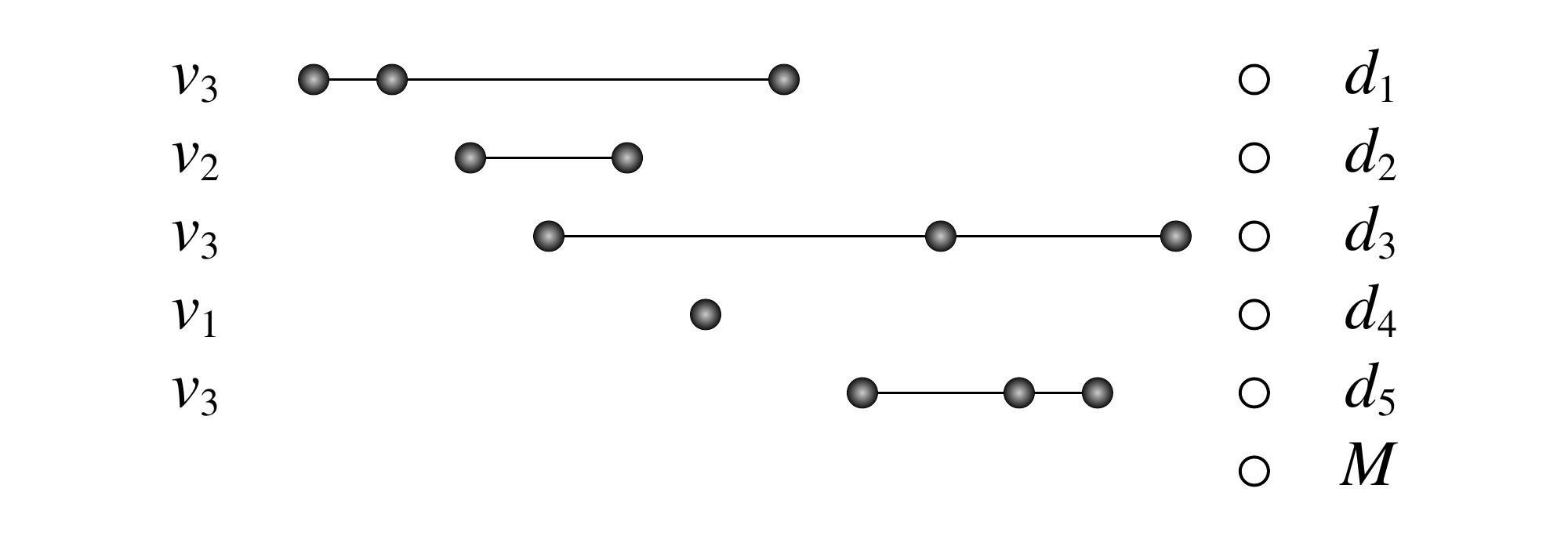}
\caption{Generating operations $d_j$, $M$ append new element at one of the
vacancies marked by empty circles. A block of size $l$ is associated with the
variable $v_l$.}
\label{fig:P-djM}
\end{figure}

The problem of construction of all set partitions $\Pi_n$ admits the following
solution. Let us consider the operations
\[
 d_j:\Pi_{n,k}\to\Pi_{n+1,k},~~ 1\le j\le k,\qquad
 M:\Pi_{n,k}\to\Pi_{n+1,k+1}
\]
defined as adding of a new element either to one of existing blocks of the
partition $\pi=\{\pi_1,\dots,\pi_k\}$, or as a new singleton:
\begin{equation}\label{SP.djM}
 d_j\pi=\{\pi_1,\dots,\pi_j\cup\{n+1\},\dots,\pi_k\},\quad
 M\pi=\{\pi_1,\dots,\pi_k,\{n+1\}\}.
\end{equation}
These operations generate partitions in the canonical form automatically, if
the element $n+1$ (or the set $\{n+1\}$) is appended to the corresponding list,
see fig.~\ref{fig:P-djM}, where the column on the right contains all possible
vacancies for the new element.

\begin{statement}\label{st:SP.dM}
Any partition of $[n]$ is generated, in a unique way, by operations $d_j,M$
applied to the seed partition $\varnothing$.
\end{statement}
\begin{proof}
The sequence of operations is recovered uniquely by deleting the elements in
the reverse order, from $n$ to $1$.
\end{proof}

In terms of the restricted growth sequences $s=(s_1,\dots,s_n)\in R_n$, the
generating operations amount to appending of an integer $s_{n+1}\in[k+1]$,
where $k=\max s$, and the operation $M$ corresponds to the value $k+1$. In
fact, $s$ encodes the sequence of operations which generate a given partition.
Alternatively, it can be written as follows. Let $\Phi_n$ denote the set of
words consisting of $n$ characters $d_j$ and $M$, such that the subscript for
any character $d$ may take values in the range from 1 to the number of
occurrences of $M$ to the right from this character. Statement \ref{st:SP.dM}
establishes the bijection $\varphi\mapsto\varphi(\varnothing)$ between $\Phi_n$
and $\Pi_n$. Although we do not use this representation of set partitions in
this section, its generalizations will be important in the study of special
partition in sections \ref{s:NOP}, \ref{s:AP}.

The generating operations provide a plain and obvious construction algorithm
for the set partitions, which is implemented, in particular, in the {\em
Combinatorica} package \cite{Pemmaraju_Skiena_2003}. It should be mentioned,
that there exist also more effective algorithms based on the Gray codes
\cite[Ch.7.2.1.5]{Knuth_2011}, \cite{Mansour_2013}.

Let us turn to the problem of counting of the set partitions. Let the variable
$v_l$ be assigned to any block $\pi_j$ with $l$ elements. In the corresponding
restricted growth sequence, $l$ is the multiplicity of the entry $j$. A
partition is associated with the monomial equal to the product of variables
assigned to all blocks:
\begin{equation}\label{ppi}
 p(\pi)=\prod^{|\pi|}_{j=1}v_{|\pi_j|}.
\end{equation}
Summation over all partitions of $[n]$ defines the ({\em complete exponential})
{\em Bell polynomials} (see e.g. \cite{Comtet_1974})
\begin{equation}\label{Yn}
 Y_n(v_1,\dots,v_n)
  =\sum_{\pi\in\Pi_n}\prod^{|\pi|}_{j=1}v_{|\pi_j|}
  =\sum_{s\in R_n}\prod^{\max s}_{j=1}v_{|\{i:\:s_i=j\}|}.
\end{equation}
In the polynomial $Y_n$, the degrees in any monomial $v^{k_1}_1\dots v^{k_r}_r$
are related by equality $n=k_1+2k_2+\dots+rk_r$ (that is, the polynomial is
homogeneous with respect to the weight $w(v_l)=l$), while the coefficient of
the monomial is equal to the number of partitions of $[n]$ into $k_1$ blocks
with 1 element, \dots, $k_r$ blocks with $r$ elements, for instance:
\[
\setlength{\arraycolsep}{3pt}
\begin{array}[t]{rc}
 Y_1=& v_1 \\
   n=& 1 \\[5pt]
     & 1
\end{array}\qquad
\begin{array}[t]{rccc}
 Y_2=& v_2 &+& v^2_1 \\
   n=&  2  &=& 1+1   \\[5pt]
     & 12  & & 1|2
\end{array}\qquad
\begin{array}[t]{rccccc}
 Y_3=& v_3 &+& 3v_1v_2 &+& v^3_1 \\
   n=&   3 &=& 1+2     &=& 1+1+1 \\[5pt]
     & 123 & & 1|23    & & 1|2|3 \\
     &     & & 12|3    & &       \\
     &     & & 13|2    & &
\end{array}
\]\[
\setlength{\arraycolsep}{3pt}
\begin{array}[t]{rccccccccc}
 Y_4=&v_4  &+& 4v_1v_3 &+& 3v^2_2 &+& 6v^2_1v_2 &+& v^4_1   \\
   n=&4    &=& 1+3     &=& 2+2    &=& 1+1+2     &=& 1+1+1+1 \\[5pt]
     &1234 & & 1|234   & & 12|34  & & 1|2|34    & & 1|2|3|4 \\
     &     & & 123|4   & & 13|24  & & 1|23|4    & &         \\
     &     & & 124|3   & & 14|23  & & 1|24|3    & &         \\
     &     & & 134|2   & &        & & 12|3|4    & &         \\
     &     & &         & &        & & 13|2|4    & &         \\
     &     & &         & &        & & 14|2|3    & &
\end{array}
\]
At $n=0$, we have $Y_0=p(\varnothing)=1$, in accordance with the common
convention that a product over empty set is equal to $1$.

The number of set partitions with blocks of prescribed size is computed
explicitly and we arrive to the formula
\begin{equation}\label{Yn!}
 Y_n=\sum_{k_1+2k_2+\dots+rk_r=n}
 \frac{n!}{(1!)^{k_1}\dots(r!)^{k_r}k_1!\dots k_r!}\,v^{k_1}_1\dots v^{k_r}_r,
\end{equation}
where the sum is taken over all {\em partitions of the number} $n$. The
following statement provides a more effective computation method. In this
statement and further on, $D=\partial_x$ denotes the differentiation with
respect to the variable $x$, and variables $v_l$ are interpreted as derivatives
of a function $v=v_0=v(x)$.

\begin{statement}\label{st:Yn}
Polynomials $Y_n$ in variables $v_l=D^l(v)$ are governed by the recurrence
relation
\begin{equation}\label{Yn-rec}
 Y_0=1,\qquad Y_{n+1}=(D+v_1)(Y_n),\quad n\ge0.
\end{equation}
The exponential generating function of the sequence $Y_n$ is
\begin{equation}\label{Yn-egf}
 \sum^\infty_{n=0}Y_n\frac{z^n}{n!}
  =\exp\left(\sum^\infty_{n=1}v_n\frac{z^n}{n!}\right).
\end{equation}
\end{statement}
\begin{proof}
Let $\pi\in\Pi_n$. The action of the differentiation $D$ on the monomial
$p(\pi)$ amounts, according to the Leibniz rule, to replacement of $v_l$ by
$v_{l+1}$ for each factor in turn, taking the multiplicity into account. In the
partitions language, this corresponds to adding of a new element to one of the
blocks in turn. As the result, we obtain the sum of monomials $p(d_j\pi)$ with
all admissible values of $j$. The multiplication of monomial $p(\pi)$ by $v_1$
yields the monomial $p(M\pi)$.

Thus, $(D+v_1)(p(\pi))=\sum_{\pi'}p(\pi')$, where the sum is taken over the
partitions $\pi'$ obtained from $\pi$ by adding the element $n+1$ in all
possible ways. Summation over $\pi$ brings to equation (\ref{Yn-rec}), taking
Statement \ref{st:SP.dM} into account. This equation can be cast to the form
$Y_{n+1}=e^{-v}De^v(Y_n)$ which implies $Y_n=e^{-v}D^n(e^v)$. For the
exponential generating function, we obtain
\[
 \sum^\infty_{n=0}Y_n\frac{z^n}{n!}
 =e^{-v}\sum^\infty_{n=0}D^n(e^v)\frac{z^n}{n!}=e^{v(x+z)-v(x)}
\]
and expansion of $v(x+z)$ into the Taylor series with respect to $z$ gives
(\ref{Yn-egf}).
\end{proof}

\begin{table}[t]
\begin{align*}
 Y_0 &= 1\\
 Y_1 &= v_1\\
 Y_2 &= v_2+v^2_1\\
 Y_3 &= v_3+3v_1v_2+v^3_1\\
 Y_4 &= v_4+(4v_1v_3+3v^2_2)+6v^2_1v_2+v^4_1\\
 Y_5 &= v_5+(5v_1v_4+10v_2v_3)+(10v^2_1v_3+15v_1v^2_2)+10v^3_1v_2+v^5_1
\end{align*}
\[
 \begin{array}{c|llllllllll}
 n\backslash k &
    0& 1& 2 &   3&   4&   5&  6& 7&~& B_n\\
 \cline{1-9}
 0& 1&  &   &    &    &    &   &  && 1\\
 1& 0& 1&   &    &    &    &   &  && 1\\
 2& 0& 1&  1&    &    &    &   &  && 2\\
 3& 0& 1&  3&   1&    &    &   &  && 5\\
 4& 0& 1&  7&   6&   1&    &   &  && 15\\
 5& 0& 1& 15&  25&  10&   1&   &  && 52\\
 6& 0& 1& 31&  90&  65&  15&  1&  && 203\\
 7& 0& 1& 63& 301& 350& 140& 21& 1&& 877
 \end{array}
\]
\captionsetup{width=0.95\textwidth}
\caption{Complete exponential Bell polynomials; the terms of the same degree are
collected in parentheses. Stirling numbers of the second kind
$\StirlingII{n}{k}=|\Pi_{n,k}|$; sums over rows are Bell numbers
$B_n=|\Pi_n|$.}\label{t:Yn}
\end{table}

The few first polynomials $Y_n$ are written down in table \ref{t:Yn}. The terms
of degree $k$ called partial Bell polynomials $Y_{n,k}$ appear in the Fa\`a di
Bruno formula for the $n$-th derivative of a composite function:
\begin{equation}\label{Faa}
 D^n(f(v))=f'(v)Y_{n,1}+f''(v)Y_{n,2}+\dots+f^{(n)}(v)Y_{n,n}
\end{equation}
(see discussion in \cite{Johnson_2002}). By summing up the coefficients of
$Y_{n,k}$, which is equivalent to identifying all variables $v_l$, we forget
about the sizes of blocks and consider just their number in a given partition.
This gives rise to the Bell--Touchard polynomials in one variable
\[
 B_n(v)=Y_n(v,\dots,v)=\sum^n_{k=0}\StirlingII{n}{k}v^k,
\]
where the coefficient of $v^k$, the {\em Stirling number of the second kind},
is equal to the number of partitions of $[n]$ into $k$ blocks \citeo{A048993}.
By definition, $\StirlingII{n}{0}=0$ at $n>0$ and $\StirlingII{0}{0}=1$, in
accordance with the convention $\Pi_{0,0}=\{\varnothing\}$. The substitution
$v_l=v$ turns the recurrence relation (\ref{Yn-rec}) into
\[
 B_0(v)=1,~~ B_{n+1}(v)=(v\partial_v+v)(B_n(v))
 \quad\Leftrightarrow\quad
 \StirlingII{n+1}{k}=\StirlingII{n}{k-1}+k\StirlingII{n}{k}
\]
and the generating function (\ref{Yn-egf}) takes the form
\[
 \sum^\infty_{n=0}B_n(v)\frac{z^n}{n!}=\exp(v(e^z-1)).
\]
The total number of partitions of a set with $n$ elements, the {\em Bell} or
{\em exponential number} \citeo{A000110}, is defined by equations
\[
 B_n=B_n(1)=Y_n(1,\dots,1)=\sum^n_{k=0}\StirlingII{n}{k},\quad
 \sum^\infty_{n=0}B_n\frac{z^n}{n!}=e^{e^z-1}.
\]

\subsection{Potential Burgers hierarchy}\label{s:pot-Burgers}

Let us recall the notion of generalized symmetry from the theory of integrable
equations. Let $v_{,t}=\partial_t(v)$ and $v_i=D^i(v)$ denote, as before, the
derivatives with respect to the distinguished variable $x$. The evolution
equation
\[
 v_{,t}=f(v_0,\dots,v_m)
\]
admits the symmetry $v_{,\tau}=g(v_0,\dots,v_n)$ (classical at $n\le1$ and
generalized, or higher, at $n>1$), if the corresponding flows commute, that is,
the equality
\[
 f_*(g)=g_*(f)
\]
holds identically with respect to $v_i$, where $f_*$ is the differential
operator
\[
 f_*=\partial_{v_0}(f)+\dots+\partial_{v_m}(f)D^m.
\]
The equation is considered integrable if it admits a sequence (hierarchy) of
symmetries of arbitrarily large order. Moreover, in a typical situation, the
higher symmetries commute with each other. The potential Burgers hierarchy
\begin{equation}\label{pot-Burgers}
 v_{,t_n}=Y_n(v_1,\dots,v_n),\quad n=0,1,2,\dots
\end{equation}
satisfies this commutativity property. This follows easily from its equivalence
to the linear hierarchy
\begin{equation}\label{heat}
 \psi_{,t_n}=\psi_n,
\end{equation}
namely, the change $\psi=e^v$ results in
\[
 v_{,t_n}=\psi^{-1}D^n(\psi)=e^{-v}D^n(e^v)=(D+v_1)^n(1)=Y_n(v_1,\dots,v_n).
\]
The mutual commutativity of the flows (\ref{heat}) is obvious
($D_n(\psi_m)=D_m(\psi_n)$) and it is not difficult to prove that this property
is preserved under the point transformations, see e.g. \cite{Ibragimov_1983}.

\begin{remark}
More generally, in the theory of integrability it is common to consider
equivalent the equations related by point or contact transformations and also
some non-invertible differential substitutions, because these changes do not
affect such properties as the existence of higher symmetries, conservation
laws, Lax pairs and so on. The combinatorial interpretation is, however, not
invariant. In contrast, we have seen that meaningful combinatorics may appear
just from nothing, as a result of simple change of variables between equations
(\ref{heat}) and (\ref{pot-Burgers}). Further examples of this kind are given
in sections \ref{s:u2u2}, \ref{s:IS}. Thus, a combinatorics is associated with
a concrete form of a hierarchy.
\end{remark}

The commutativity follows also from the identity (\ref{Y*Y}) below, which
coincides with the recurrence relation (\ref{Yn-rec}) at $m=1$, $Y_1=v_1$,
$(Y_1)_*=D$. The general case can be easily proven by induction with respect to
$m$, but instead we will generalize the combinatorial proof of equation
(\ref{Yn-rec}). First, let us introduce the following notation.

The {\em concatenation} of partitions $\pi\vdash[n]$, $\rho\vdash[m]$ is the
partition
\begin{equation}\label{|}
 \pi|\rho=\pi\cup(\rho+n)\vdash[m+n],
\end{equation}
where $\rho+n$ denotes the partition of $[n+1,n+m]$ obtained from $\rho$ by
adding $n$ to all elements. (Compare with the {\em Mathematica} language
convention that adding of a scalar to a list acts on each entry:
$(a_1,\dots,a_k)+b=(a_1+b,\dots,a_k+b)$.)

\begin{figure}[t]
\gr{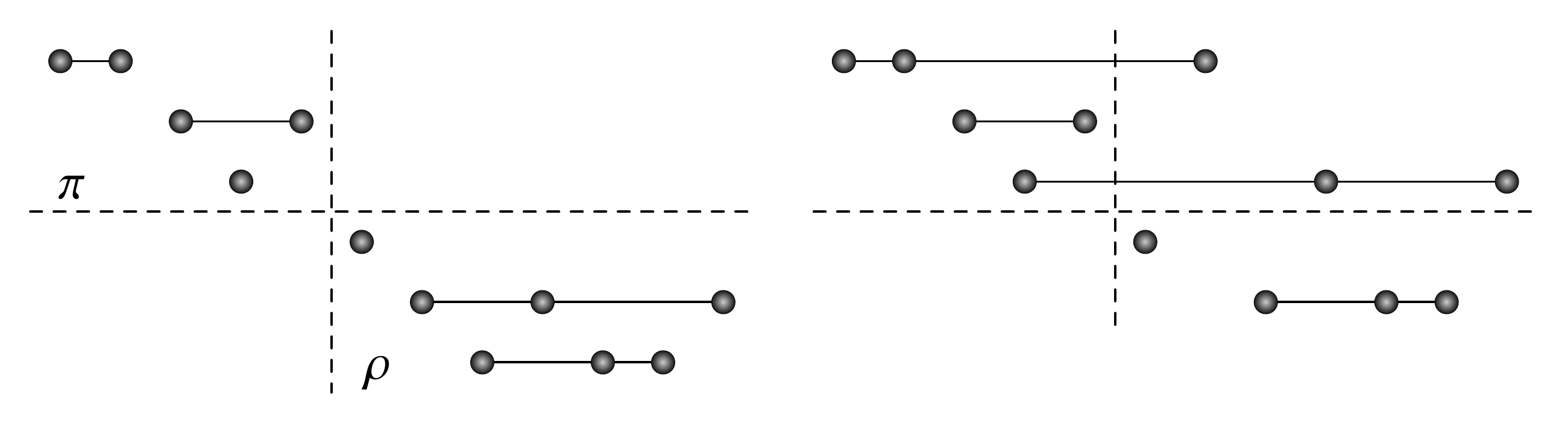}
\caption{Towards the proof of identity (\ref{Y*Y}).}
\label{fig:concatenation}
\end{figure}

\begin{statement}\label{st:YY}
The Bell polynomials satisfy the identity
\begin{equation}\label{Y*Y}
 Y_{m+n}=(Y_m)_*(Y_n)+Y_mY_n.
\end{equation}
\end{statement}
\begin{proof}
Given $\pi\vdash[n]$, $\rho\vdash[m]$, let us construct the partition
$\sigma=\pi|\rho$ (fig.~\ref{fig:concatenation} on the left). The corresponding
monomials (\ref{ppi}) satisfy the relation $p(\sigma)=p(\pi)p(\rho)$. Next, let
us consider the partitions obtained from $\sigma$ by deleting of one of the
size $l$ blocks of $\rho+n$ and adding its elements to blocks of $\pi$, in all
possible ways (one such partition is shown on the right of
fig.~\ref{fig:concatenation}). The sum of the corresponding monomials is equal
to $D^l(p(\pi))\partial_{v_l}(p(\rho))$.

Vise versa, given $\sigma\vdash[m+n]$, let us consider the blocks with the
minimal element not greater than $n$. Their intersections with $[n]$ uniquely
define the partition $\pi$, while the block formed from the rests of these
blocks and the rest blocks of $\sigma$ constitute the partition $\rho+n$.
Therefore, the above operations give all partitions $\Pi_{m+n}$, without
repetitions, when $\pi$ and $\rho$ run, respectively, over $\Pi_n$ and $\Pi_m$.
The summation of monomials gives
\[
 \sum_{\sigma\in\Pi_{m+n}}p(\sigma)=
 \sum_{\pi\in\Pi_n}\sum_{\rho\in\Pi_m}\Bigl(p(\pi)p(\rho)
  +\sum_l D^l(p(\pi))\partial_{v_l}(p(\rho))\Bigr),
\]
as required.
\end{proof}

\subsection{Burgers hierarchy}\label{s:Burgers}

The right hand sides of equations (\ref{pot-Burgers}) do not contain $v$, which
makes possible the substitution $u=u_0=v_1$. This brings to the Burgers
hierarchy
\begin{equation}\label{Burgers}
 u_{,t_n}=D(Y_n(u_0,\dots,u_{n-1})),\quad n=1,2,\dots,
\end{equation}
homogeneous with respect to the weight $w(u_j)=j+1$. The few first equations
are
\begin{align*}
 u_{,t_1} &= u_1,\\
 u_{,t_2} &= u_2+2uu_1,\\
 u_{,t_3} &= u_3+(3uu_2+3u^2_1)+3u^2u_1,\\
 u_{,t_4} &= u_4+(4uu_3+10u_1u_2)+(6u^2u_2+12uu^2_1)+4u^3u_1,\\
 u_{,t_5} &= u_5+(5uu_4+15u_1u_3+10u^2_2)+(10u^2u_3+50uu_1u_2+15u^3_1)\\
         &\qquad +(10u^3u_2+30u^2u^2_1)+5u^4u_1, \quad \dots\;.
\end{align*}
What is the combinatorial interpretation of the coefficients in this case? The
differentiation in equation (\ref{Burgers}) corresponds to adding, turn by
turn, of an element to blocks of partitions $\Pi_n$. Since we do not add a new
block in this case, hence the partitions under consideration are constructed in
the same manner as in section \ref{s:gf}, but the operation $M$ is not used on
the last step. As a result, only those of partitions $\Pi_{n+1}$ are
constructed, which do not contain element $n+1$ as a singleton; for instance,
at $n=3$:
\[
\setlength{\arraycolsep}{1pt}
\begin{array}{ccccccccccccccc}
 u_2&+&3uu_1&+&u^3&\quad\stackrel{D}{\longrightarrow}\quad
                 & u_3&+&3uu_2      &+&3u^2_1&+& 3u^2u_1  &+&0u^4     \\[3pt]
 123&&1|23&&1|2|3&&1234&&      1|234 &&12|34&&      1|2|34 &&\sout{1|2|3|4}\\
    &&12|3&&     &&    &&\sout{123|4}&&13|24&&\sout{1|23|4}&&              \\
    &&13|2&&     &&    &&      124|3 &&14|23&&      1|24|3 &&              \\
    &&    &&     &&    &&      134|2 &&     &&\sout{12|3|4}&&              \\
    &&    &&     &&    &&            &&     &&\sout{13|2|4}&&              \\
    &&    &&     &&    &&            &&     &&      14|2|3 &&
\end{array}
\]
Renumbering makes possible to replace the element $n+1$ with any other one, and
we arrive to the following statement.

\begin{statement}\label{st:Burgers}
In the Burgers hierarchy, the coefficient of the monomial $u^{k_0}_0\dots
u^{k_r}_r$ is equal to the number of partitions of the set
$[k_0+2k_1+\dots+(r+1)k_r]$ into $k_0$ blocks with 1 element, \dots, $k_r$
blocks with $r+1$ elements, under the additional condition that a distinguished
element of the set is not a singleton.
\end{statement}

The total number of partitions of $[n+1]$ such that a distinguished element is
not a singleton is found by setting all variables equal to $1$:
\begin{multline*}\qquad
 D(Y_n(u_0,\dots,u_{n-1}))|_{u_j=1}=B'_n(1)=\\
 =\sum^n_{k=0}k\StirlingII{n}{k}
 =\sum^{n+1}_{k=0}\StirlingII{n+1}{k}-\sum^n_{k=0}\StirlingII{n}{k}
 =B_{n+1}-B_n, \quad n\ge0.\qquad
\end{multline*}
These integers form the sequence
\[
 0,~ 1,~ 3,~ 10,~ 37,~ 151,~ 674,~ 3263,~ 17007,~ 94828,~ 562595,~ \dots\;.
\]
According to \citeo{A005493}, it can be characterized also as the total number
of blocks in all partitions of $[n]$. Indeed, the partitions of $[n+1]$ under
consideration are obtained from partitions of $[n]$ by enlarging of one of the
blocks, and this operation is applied exactly as many times as many blocks
there are in all partitions.

\section{Natural growth sequences}\label{s:n-seq}

\subsection{Statistics}\label{s:seq_stat}

\begin{figure}[t]
\gr{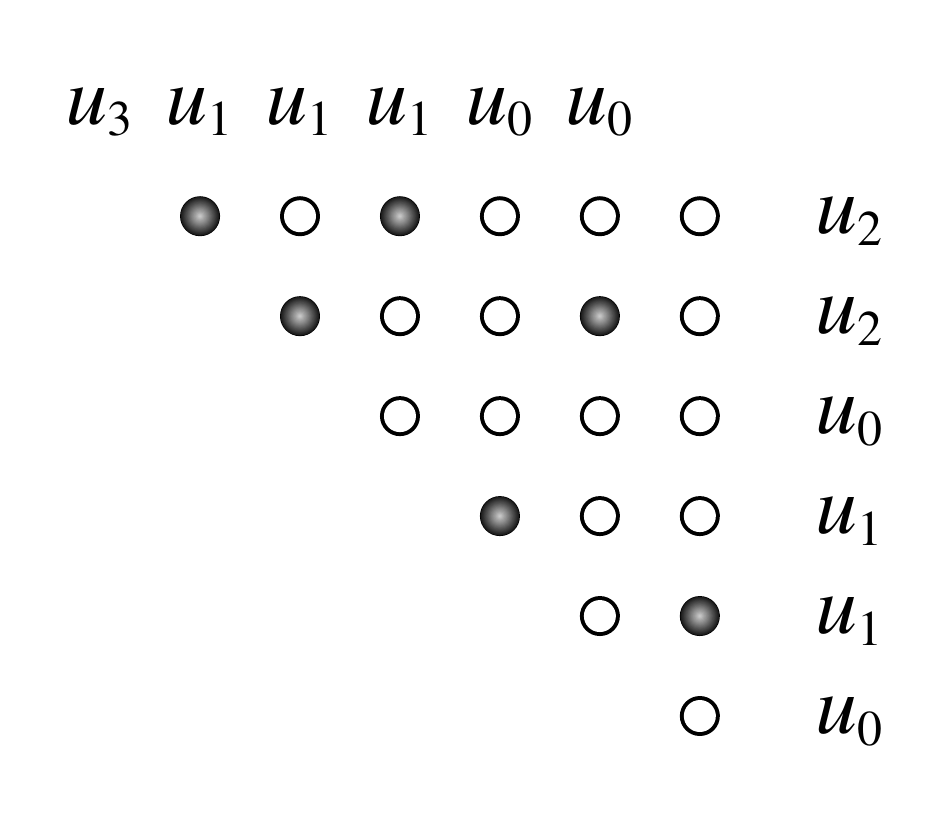}
\caption{Sequence $a=(1,2,1,4,2,5)$. Distribution by rows gives
$p(a)=u^2u^2_1u^2_2$, distribution by diagonals gives $\bar p(a)=u^2u^3_1u_3$.}
\label{fig:n-seq}
\end{figure}

Let $T_n$ denote the set of integer sequences $a=(a_1,\dots,a_n)$ satisfying
the {\em natural growth} condition $1\le a_i\le i$. The sequence $a$ can be
matched to an upper triangular $n\times n$ matrix, with one unit in each column
and rest elements equal to zero, see fig.~\ref{fig:n-seq}. It is clear that
$T_n=[1]\times[2]\times\cdots\times[n]$, that is, the structure of this set is
simpler comparing to the set $R_n$ of the restricted growth sequences
(\ref{rgs}). In particular, $|T_n|=n!$, instead of the Bell numbers for
$|R_n|$. Nevertheless, these sequences also support a quite meaningful
generating function.

For any sequence $a\in T_n$, let us count the occurrences of integers from $1$
to $n$ and form the monomial
\begin{equation}\label{pa}
 p(a)=u_{|\{i:\:a_i=1\}|}\cdots u_{|\{i:\:a_i=n\}|}.
\end{equation}
Summation over all sequences defines the polynomials
\begin{equation}\label{hn}
 h_n(u_0,\dots,u_n)
  =\sum_{a\in T_n}\prod^n_{j=1}u_{|\{i:\:a_i=j\}|},
\end{equation}
analogous to the definition of the Bell polynomials (\ref{Yn}) for the
restricted growth sequences. In both cases, the definition of monomials is
related with counting of elements in the level sets, which are the rows of the
upper triangular matrix or the partition blocks. Notice, that one can count the
units in the diagonals of the matrix instead of the rows, resulting in the
monomial $\bar p(a)=\prod^n_{j=1}u_{|\{i:\:a_i=i-j+1\}|}$. In general, $\bar
p(a)\ne p(a)$, but, nevertheless, the summation over all $a$ gives the same
result. Indeed, if we reverse a part of each column, from the first element to
the diagonal one, then a matrix of the same type appears, and this transform is
bijective and takes the rows into the diagonals. For the sequences, it is given
by the map $a\to(2,\dots,n+1)-a$.

In order to derive the recurrence relation for $h_n$ we assume, as before, that
$u=u_0=u(x)$, $D=\partial_x$.

\begin{statement}\label{st:hn}
The polynomials $h_n$ in variables $u_l=D^l(u)$ satisfy the recurrence relation
\begin{equation}\label{hn-rec}
 h_1=u_1,\quad h_{n+1}=D(uh_n),\quad n=1,2,\dots,
\end{equation}
that is, $h_n=(Du)^n(1)$.
\end{statement}
\begin{proof}
Let us interpret the sequence $a$ as a distribution of enumerated balls over
enumerated boxes, possibly empty, such that $i$-th ball is allowed to be only
in first $i$ boxes. The variable $u_l$ is assigned to each box with $l$ balls,
and the monomial (\ref{pa}) is equal to their product. Generating operations
for $T_n$ are extremely simple: the passage to $T_{n+1}$ amounts to appending
of an arbitrary integer $a_{n+1}\in[n+1]$. Let $d_ja=(a_1,\dots,a_n,j)$,
$j=1,\dots,n+1$. These operations are interpreted as follows: first, we add the
empty box number $n+1$, this multiplies $p(a)$ by $u=u_0$; next, we put the
ball number $n+1$ into the box number $j$, this replaces the variable $u_l$
assigned to this box with $u_{l+1}$. As a result, $p(a)$ is mapped into the sum
of monomials over all sequences $a'$ with the same first $n$ elements as in
$a$:
\[
 \sum_{a'}p(a')=\sum^{n+1}_{j=1}p(d_ja)
  =\sum^n_{l=0}u_{l+1}\partial_{u_l}(up(a))=D(up(a)).
\]
Summation over $a\in T_n$ completes the proof.
\end{proof}

In contrast to the Bell polynomials, $h_n$ are homogeneous not only with
respect to the weight $w(u_l)=l$, but also with respect to the degree. The few
first polynomials are written down in table \ref{t:hn}, with the terms
collected by powers of the distinguished variable $u=u_0$. Let the rest
variables be equal to $v$ and denote $h_n(u,v,\dots,v)=vE_n(u,v)$, this gives
the homogeneous polynomials in two variables
\[
 E_n(u,v)=\sum^n_{k=1}\Euler{n}{k}u^{n-k}v^{k-1}
  =u^{n-1}+\dots+v^{n-1},
\]
where the {\em Euler number} $\Euler{n}{k}$ is equal to the number of sequences
from $T_n$ which take $k$ different values \citeo{A008292}. The total sum of
the coefficients is, obviously, $E_n(1,1)=|T_n|=n!$.

\begin{table}[t]
\begin{align*}
 h_1 &= u_1\\
 h_2 &= uu_2+u^2_1\\
 h_3 &= u^2u_3+4uu_1u_2+u^3_1\\
 h_4 &= u^3u_4+u^2(7u_1u_3+4u^2_2)+11uu^2_1u_2+u^4_1\\
 h_5 &= u^4u_5+u^3(11u_1u_4+15u_2u_3)
         +u^2(32u^2_1u_3+34u_1u^2_2)+26uu^3_1u_2+u^5_1
\end{align*}
\[
 \begin{array}{c|lllllll}
 n\backslash k &
    1& 2  & 3   & 4   & 5   & 6  & 7\\
 \hline
 1& 1&    &     &     &     &    &  \\
 2& 1& 1  &     &     &     &    &  \\
 3& 1& 4  & 1   &     &     &    &  \\
 4& 1& 11 & 11  & 1   &     &    &  \\
 5& 1& 26 & 66  & 26  & 1   &    &  \\
 6& 1& 57 & 302 & 302 & 57  & 1  &  \\
 7& 1& 120& 1191& 2416& 1191& 120& 1
 \end{array}
\]
\caption{Polynomials $h_n$. Euler numbers $\Euler{n}{k}$.}\label{t:hn}
\end{table}

The above substitution replaces the operator $D$ with
$v(\partial_u+\partial_v)$ and the recurrence relation (\ref{hn-rec}) takes the
form
\[
 E_1=1,\quad E_{n+1}=(\partial_u+\partial_v)(uvE_n),\quad n=1,2,\dots,
\]
which implies the symmetry property and the recurrence relation for the Euler
numbers
\[
 \Euler{n}{k}=\Euler{n}{n+1-k},\quad
 \Euler{n}{k}=(n+1-k)\Euler{n-1}{k-1}+k\Euler{n-1}{k}.
\]

\begin{remark}
It should be mentioned, that the Euler numbers admit also many other
interpretations, related with the permutations of $n$ elements, in particular,
$\Euler{n}{k}$ is equal to the number of permutations
$\sigma=(\sigma_1,\dots,\sigma_n)\in S_n$ with $k-1$ descents (that is,
positions $i$, such that $\sigma_i>\sigma_{i+1}$), and also to the number of
permutations with $k-1$ exceedances (positions $i$, such that $\sigma_i>i$). In
fact, the permutations admit a wide variety of combinatorial interpretations
and representations, resulting in diversity of associated statistics, see e.g.
\cite{Stanley_1986,Bona_2004}. The natural growth sequences also appears in the
theory of permutations, as the inversion vectors and Lehmer codes. Recall, that
the inversion vector is a sequence $b=(b_1,\dots,b_n)$, such that $b_i$ is
equal to the number of elements $\sigma$ on the left of $i$ and greater than
$i$, that is $b_i=|\{j:j<(\sigma^{-1})_i\wedge\sigma_j>i\}|$. It is clear that
$b_i\in[0,n-i]$, therefore, adding 1 to the entries of an inversion vector and
taking its reverse bring to a sequence from $T_n$. This mapping $S_n\to T_n$ is
a bijection, see e.g. \cite{Stanley_1986, Pemmaraju_Skiena_2003}.
\end{remark}

\subsection{Hierarchy of equation $u_t=u^2u_2$}\label{s:u2u2}

The polynomials $h_n$ appear, up to some denotation, in the computation of the
inverse function derivatives, like the Bell polynomials appear in the formula
for the derivatives of a composite function (\ref{Faa}). Indeed, let
$u=1/D(v(x))$, $D=d/dx$, then
\[
 \frac{d^nx}{dv^n}=(uD)^n(x)=u(Du)^{n-1}(1)=uh_{n-1}(u,\dots,u_{n-1})
\]
(cf. with the definition of $h_n$ in \citeo{A145271}). Essentially, this
formula is used in the hodograph type transformation of the linear hierarchy
(\ref{heat}) which we rewrite now in the form
\[
 \partial_{\tilde t_n}(\tilde v)=\partial^n_{\tilde x}(\tilde v).
\]
Under the change of variables $v=\tilde x$, $x=\tilde v$, $t_n=\tilde t_n$, the
derivatives are computed according to the rule
\[
 \partial_{\tilde x}=\frac{1}{v_1}D,\quad
 \partial_{\tilde t_n}=\partial_{t_n}-\frac{v_{,t_n}}{v_1}D,\quad
 D=\partial_x,\quad v_1=D(v)
\]
and the equation takes the form
$v_{,t_n}=-\bigl(D\frac{1}{v_1}\bigr)^{n-1}(1)$. Further differential
substitution $u=1/v_1$ brings to equations
\begin{equation}\label{unun}
 u_{,t_n}=u^2D(Du)^{n-1}(1)=u^2D(h_{n-1}(u,\dots,u_{n-1})),\quad n=2,3,\dots,
\end{equation}
in particular, first two flows are
\[
 u_{,t_2}=u^2u_2,\quad u_{,t_3}=u^3u_3+3u^2u_1u_2
\]
(a discussion of this example can be found in
\cite[sect.20.1]{Ibragimov_1983}).

The commutativity of equations (\ref{unun}) follows, like in section
\ref{s:pot-Burgers}, either from their relation to equations (\ref{heat}) or
intermediately from the identity
\[
 h_{m+n+1}=(h_m)_*u^2D(h_n)+D(uh_mh_n),
\]
which can be easily proved by induction with respect to $m$. It would be
interesting to find its combinatorial meaning, like in Statement \ref{st:YY}.

\section{$B$ type set partitions}\label{s:B}

\subsection{Generating operations and statistics}\label{s:B-stat}

Set partitions of $B$ type \cite{Dowling_1973}, see also \cite{Benoumhani_1996,
Reiner_1997, Suter_2000}, appear if we employ the reflection symmetry of the
set under consideration.

\begin{definition}
A partition $\pi$ of the set $\{-n,\dots,n\}$ is called $B$ type partition if:

1) $\pi=-\pi$, that is, for any block $\beta\in\pi$ also $-\beta\in\pi$;

2) $\pi$ contains just one block $\pi_0$ (called $0$-block), such that
$\pi_0=-\pi_0$.
\end{definition}

\begin{figure}[t]
\gr{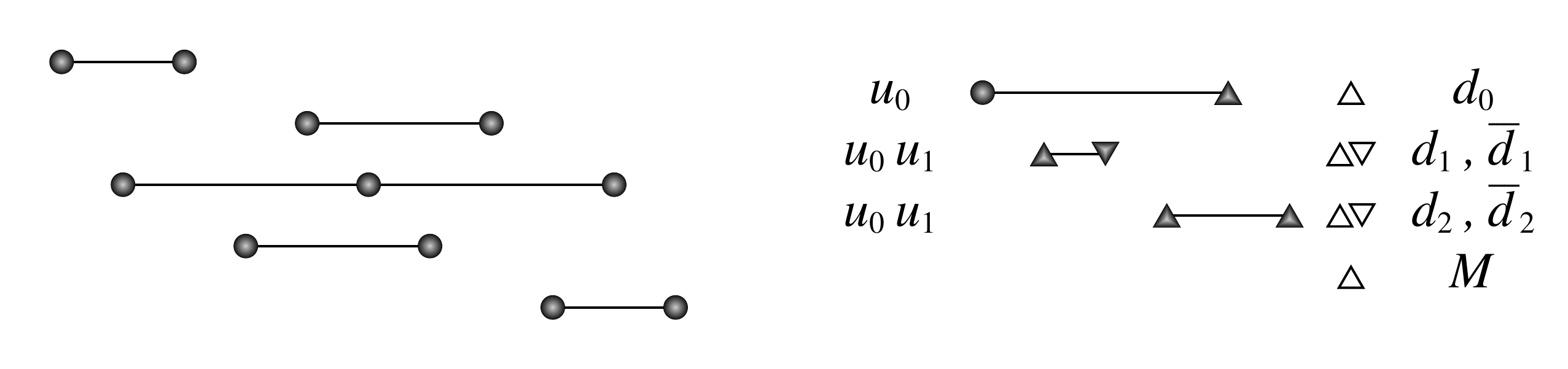}
\captionsetup{width=0.95\textwidth}
\caption{On the left: full representation of a $B$ type partition;
the blocks are sorted downwards by increase of elements with minimal absolute
value. On the right: representation of the canonical form; negative elements
are marked {\color[gray]{0.3}$\blacktriangledown$}, positive
{\color[gray]{0.3}$\blacktriangle$}. Generating operations add element to one
of the vacant positions marked by empty triangles. The corresponding monomials
(\ref{Bppi}) are written near to the blocks.}
\label{fig:BP-djM}
\end{figure}

All such partitions are denoted $\Pi^B_n$, and partitions with $k$ block pairs
are denoted $\Pi^B_{n,k}$, $0\le k\le n$. The canonical form of $B$ type
partition is defined as follows: in $0$-block, all negative elements are
removed; in all blocks, the elements are sorted by increase of the absolute
value; from any block pair we keep only the block with positive first element;
the blocks are sorted by increase of their first elements. Moreover, it is
convenient to write $\bar j$ instead of $-j$. For instance, the partition
\[
 \{\{-5,-3\},\{-1,2\},\{-4,0,4\},\{-2,1\},\{3,5\}\}
\]
is encoded as $04|1\bar2|35$. The pictorial representation is clear from
fig.~\ref{fig:BP-djM}.

Notice, that deleting elements $\pm n$ from any partition $\Pi^B_n$ yields a
partition from $\Pi^B_{n-1}$. Therefore, $\Pi^B_n$ is obtained from
$\Pi^B_{n-1}$ by adding $\pm n$ in all possible ways, namely:
\begin{list}{}{}
\item $d_0:\Pi^B_{n-1,k}\to\Pi^B_{n,k}$,
      adding both elements $\pm n$ to 0-block;
\item $d_j:\Pi^B_{n-1,k}\to\Pi^B_{n,k}$,
      adding $\pm n$ to the blocks $\pm\pi_j$, $j=1,\dots,k$;
\item $\bar d_j:\Pi^B_{n-1,k}\to\Pi^B_{n,k}$,
      adding $\pm n$ to the blocks $\mp\pi_j$, $j=1,\dots,k$;
\item $M:\Pi^B_{n-1,k}\to\Pi^B_{n,k+1}$,
      adding a new block pair $\{-n\}$, $\{n\}$.
\end{list}
These operations generate, in a unique way, any $B$ type partition starting
from the trivial partition of the set $\{0\}$.

Let us define the mapping $p$ from $\Pi^B_n$ into the set of monomials of
variables $u_j$. Given a block $\beta$, let $\pos{\beta}$ denote the number of
its positive elements:
\[
 \pos{\beta}=|\{i\in\beta:\:i>0\}|.
\]
It is clear that the number of negative elements in the block is equal to
$\pos{-\beta}$. Let a partition $\pi\in\Pi^B_{n,k}$ consists of 0-block $\pi_0$
and block pairs $\pi_1,-\pi_1,$ \dots, $\pi_k,-\pi_k$, such that the elements
with minimal absolute value are positive in blocks $\pi_j$. We assign the
monomial of degree $2k+1$
\begin{equation}\label{Bppi}
 p(\pi)=u_{\pos{\pi_0}}u_{\pos{\pi_1}-1}u_{\pos{-\pi_1}}\cdots
  u_{\pos{\pi_k}-1}u_{\pos{-\pi_k}}
\end{equation}
to such a partition, so that all partitions $\Pi^B_n$ are associated with the
polynomial
\begin{equation}\label{an}
 a_n(u_0,u_1,\dots,u_n)= \sum_{\pi\in\Pi^B_n}u_{\pos{\pi_0}}
 \prod^{(|\pi|-1)/2}_{j=1}u_{\pos{\pi_j}-1}u_{\pos{-\pi_j}}
\end{equation}
which serves as $\mathbb Z_2$-analog of the Bell polynomial $Y_n$. As an
example, let us write down all partitions $\Pi^B_3$, collecting together the
partitions which correspond to one and the same monomial (as usual, $u=u_0$):
\[
\setlength{\arraycolsep}{1pt}
\begin{array}[t]{rccccccccc}
 a_3=& u_3  &+& 5u^2u_2       &+& 8uu^2_1   &+& 9u^4u_1    &+& u^7     \\[5pt]
     & 0123 & & 0|123         & & 0|12\bar3 & & 0|12|3     & & 0|1|2|3 \\
     &      & & 0|1\bar2\bar3 & & 0|1\bar23 & & 0|1\bar2|3 & &         \\
     &      & & 012|3         & & 01|23     & & 0|13|2     & &         \\
     &      & & 013|2         & & 01|2\bar3 & & 0|1\bar3|2 & &         \\
     &      & & 023|1         & & 02|13     & & 0|23|1     & &         \\
     &      & &               & & 02|1\bar3 & & 0|2\bar3|1 & &         \\
     &      & &               & & 03|12     & & 01|2|3     & &         \\
     &      & &               & & 03|1\bar2 & & 02|1|3     & &         \\
     &      & &               & &           & & 03|1|2     & &
\end{array}
\]

\begin{statement}\label{st:an}
Polynomials (\ref{an}) in variables $u_i=D^i(u)$, $D=\partial_x$, satisfy the
recurrence relation
\begin{equation}\label{an-rec}
 a_0=u,\quad a_{n+1}=(D+u^2)(a_n),\quad n\ge0.
\end{equation}
\end{statement}
\begin{proof}
Let $\pi\in\Pi^B_{n-1,k}$ and let us track the monomial $p(\pi)$ under the
action of generating operations:
\begin{list}{}{}
\item $d_0$: the factor $u_{\pos{\pi_0}}$ is replaced with $u_{\pos{\pi_0}+1}$;
\item $d_j$: the factor $u_{\pos{\pi_j}-1}$ is replaced with $u_{\pos{\pi_j}}$;
\item $\bar d_j$: the factor $u_{\pos{-\pi_j}}$ is replaced with $u_{\pos{-\pi_j}+1}$;
\item $M$: two factors $u$ are added.
\end{list}
Therefore, the application of all possible operations $\pi\to\pi'$ replaces the
monomial $p(\pi)$ with $\sum_{\pi'}p(\pi')=(D+u^2)(p(\pi))$; the summation over
$\pi$ completes the proof.
\end{proof}

\begin{table}[t]
\begin{align*}
 a_0&= u \\
 a_1&= u_1+u^3\\
 a_2&= u_2+4u^2u_1+u^5\\
 a_3&= u_3+(5u^2u_2+8uu^2_1)+9u^4u_1+u^7\\
 a_4&= u_4+(6u^2u_3+26uu_1u_2+8u^3_1)+(14u^4u_2+44u^3u^2_1)+16u^6u_1+u^9\\
 a_5&= u_5+(7u^2u_4+38uu_1u_3+26uu^2_2+50u^2_1u_2)+(20u^4u_3+170u^3u_1u_2\\
    &\qquad +140u^2u^3_1)+(30u^6u_2+140u^5u^2_1)+25u^8u_1+u^{11}
\end{align*}
\[
 \begin{array}{c|llllllllll}
 n\backslash k &
     0 & 1   & 2   & 3    & 4   & 5  & 6 & 7&~& \\
 \cline{1-9}
 0 & 1 &     &     &      &     &    &   &  && 1\\
 1 & 1 & 1   &     &      &     &    &   &  && 2\\
 2 & 1 & 4   & 1   &      &     &    &   &  && 6\\
 3 & 1 & 13  & 9   & 1    &     &    &   &  && 24\\
 4 & 1 & 40  & 58  & 16   & 1   &    &   &  && 116\\
 5 & 1 & 121 & 330 & 170  & 25  & 1  &   &  && 648\\
 6 & 1 & 364 & 1771& 1520 & 395 & 36 & 1 &  && 4088\\
 7 & 1 & 1093& 9219& 12411& 5075& 791& 49& 1&& 28640
 \end{array}
\]
\captionsetup{width=0.95\textwidth}
\caption{Polynomials $a_n$ are homogeneous with respect to the
weight $w(u_j)=2j+1$; parentheses separate the terms of the same degree.
$B$-analogs of the Stirling numbers of the 2nd kind $|\Pi^B_{n,k}|$; sums in
rows are equal to the Dowling numbers $|\Pi^B_n|$.}
\label{t:an}
\end{table}

The few first polynomials $a_n$ are given in table \ref{t:an}. Let us simplify
the statistics by merging the terms of the same degree; this brings to the
polynomials $a_n(u,\dots,u)=(u\partial_u+u^2)^n(u)$. The coefficients of
$u^{2k+1}$ are the numbers of partitions $\Pi^B_{n,k}$, that is, the analogs of
the Stirling numbers of the 2nd kind for $B$ type partitions \citeo{A039755}.
The total coefficient sums $a_n(1,\dots,1)$ give the total numbers of
partitions $\Pi^B_n$ which constitute the sequence of $B$-analogs of the Bell
numbers, or the {\em Dowling numbers} \citeo{A007405}.

\subsection{Ibragimov--Shabat hierarchy}\label{s:IS}

The diagram
\[
\begin{array}{lcl}
 \psi_{,t_3}=\psi_3 & &
 u_{,t_3}=u_3+3u^2u_2+9uu^2_1+3u^4u_1\\[7pt]
 \qquad\updownarrow \psi^2=s && \qquad\updownarrow u^2=v \\[7pt]
 s_{,t_3}=D\Bigl(s_2-\dfrac{3s^2_1}{4s}\Bigr) &&
 v_{,t_3}=D\Bigl(v_2-\dfrac{3v^2_1}{4v}+3vv_1+v^3\Bigr)\\[7pt]
 \qquad\uparrow s=q_1 && \qquad\uparrow v=w_1 \\[7pt]
 q_{,t_3}=q_3-\dfrac{3q^2_2}{4q_1} &
  \overset{q=e^{2w}}{\leftarrow\!\!\!-\!\!-\!\!\!\rightarrow} &
 w_{,t_3}=w_3-\dfrac{3w^2_2}{4w_1}+3w_1w_2+w^3_1
\end{array}
\]
describes the chain of point transformations and substitutions of introducing a
potential type which linearizes the Ibragimov--Shabat equation
\cite{Ibragimov_Shabat_1980}. The whole hierarchy consists of equations related
by the same transformations with equations $\psi_{,t_{2m+1}}=\psi_{2m+1}$. The
absence of even-order flows is explained by the fact that, although the changes
look quite harmless, they partially destroy the symmetry algebra. For
$\partial_{t_{2m+1}}$, the change $\psi^2=s$ brings to an equation with the
total derivative in the right hand side:
\begin{equation}\label{stn}
 s_{,t_{2m+1}}=2\psi\psi_{2m+1}
  =D\bigl(2\psi\psi_{2m}-2\psi_1\psi_{2m-1}+2\psi_2\psi_{2m-2}
   +\dots\pm\psi^2_m\bigr).
\end{equation}
In contrast, the analogous equation for $s_{,t_{2m}}$ contains the term
$\psi^2_m$ outside the parentheses, that is $s_{,t_{2m}}\not\in\mathop{\rm
Im}D$; so that the further substitution $s=q_1$ leads out of the class of
evolution equations. The structure of odd flows is described by the following
statement.

\begin{statement}
Denote $D_t=\partial_{t_1}+z^2\partial_{t_3}+z^4\partial_{t_5}+\dots$,
$A=A(z)=a_0+a_1z+a_2z^2+\dots$, $\bar A=A(-z)$, then the Ibragimov--Shabat
hierarchy is equivalent to equations
\begin{gather}
\label{ut}
 D_t(u)=\frac{1}{2u}D(A\bar A)=\frac{1}{2z}(A-\bar A)-uA\bar A,\\
\label{A}
 z(D+u^2)(A)=A-u,
\end{gather}
and the coefficients $a_n$ coincide with the polynomials (\ref{an-rec}).
\end{statement}
\begin{proof}
Let us consider the generating function
\[
 \Psi=\psi+\psi_1z+\psi_2z^2+\dots
\]
and define $A$ by equation $\Psi=\sqrt{2}e^wA$. The change of variables
$\psi=\sqrt{q_1}=\sqrt{2e^{2w}w_1}=\sqrt{2}e^wu$ maps the identity
$zD(\Psi)=\Psi-\psi$ into equation (\ref{A}), which is exactly equivalent to
recurrence relations (\ref{an-rec}).

Next, let $\bar\Psi=\Psi(-z)$, then (cf. with (\ref{stn}))
\begin{multline*}
 D(\Psi\bar\Psi)=z^{-1}(\Psi-\psi)\bar\Psi-z^{-1}\Psi(\bar\Psi-\psi)\\
  =z^{-1}\psi(\Psi-\bar\Psi)
  =2\psi(\psi_1+\psi_3z^2+\dots)=2\psi D_t(\psi)=D_t(s).
\end{multline*}
By applying $D^{-1}$, we obtain $\Psi\bar\Psi=D_t(q)=2e^{2w}D_t(w)$ which
implies
\[
 2uD_t(u)=D_t(v)=DD_t(w)=\frac{1}{2}D(e^{-2w}\Psi\bar\Psi)=D(A\bar A).
\]
The second equality in (\ref{ut}) is obtained by eliminating of derivatives in
virtue of (\ref{A}).
\end{proof}

\section{Non-overlapping partitions}\label{s:NOP}

\subsection{Generating operations}\label{s:NOP-gen}

The definition of non-overlapping partitions \cite{Flajolet_Schott_1990} makes
use of the ordering of the set $[n]$.

\begin{definition}\label{def:NOP}
Blocks $\alpha$ and $\beta$ of a partition $\pi\vdash[n]$ overlap, if
\[
 \min\alpha<\min\beta<\max\alpha<\max\beta.
\]
A partition is called {\em non-overlapping} if no pair of its blocks overlap.
Equivalently, the supports of any two blocks either do not intersect or one of
them contains another one.
\end{definition}

For instance, the partition in the fig.~\ref{fig:P-djM} does not satisfy this
requirement, because the blocks $\pi_1,\pi_3$ overlap (as well as
$\pi_2,\pi_3$; the other pairs do not).

All non-overlapping partitions of $[n]$ are denoted $\Pi^*_n$, and those which
contain $k$ multiplets (and any number of singletons) will be denoted
$\Pi^*_{n,k}$, $0\le k\le n/2$. It is natural to consider $\varnothing$
non-overlapping partition, that is, $\Pi^*_0=\Pi^*_{0,0}=\{\varnothing\}$.
Also, let $\widetilde\Pi^*_n$ denote the set of non-overlapping partitions of
$[n]$ without singletons.

\begin{remark}
In papers \cite{Flajolet_Schott_1990, Claesson_2001}, $S^*_{n,k}$ denotes the
number of non-overlapping partitions of $[n]$ with $k$ blocks of any size.
Counting multiplets only turns out to be more convenient in our construction
method for partitions. In this approach, one may collect, in the mind's eye,
all singletons of a partition into one distinguished block (possibly, empty).
Notice, that this establishes a bijection between $\Pi^*_n$ and a subset of
$\widetilde\Pi^*_{n+2}$ which consists of partitions with elements $1$ and
$n+2$ in one block.
\end{remark}

The generating operations for non-overlapping partitions are more complicated
than in the previous sections. These include the operations
\[
 d_j:\Pi^*_{n,k}\to\Pi^*_{n+1,k},~~ 0\le j\le k,\qquad
 P:\Pi^*_{n,k}\times\Pi^*_{m,l}\to\Pi^*_{n+m+2,k+l+1}
\]
which we now describe.

\begin{figure}[t]
\gr{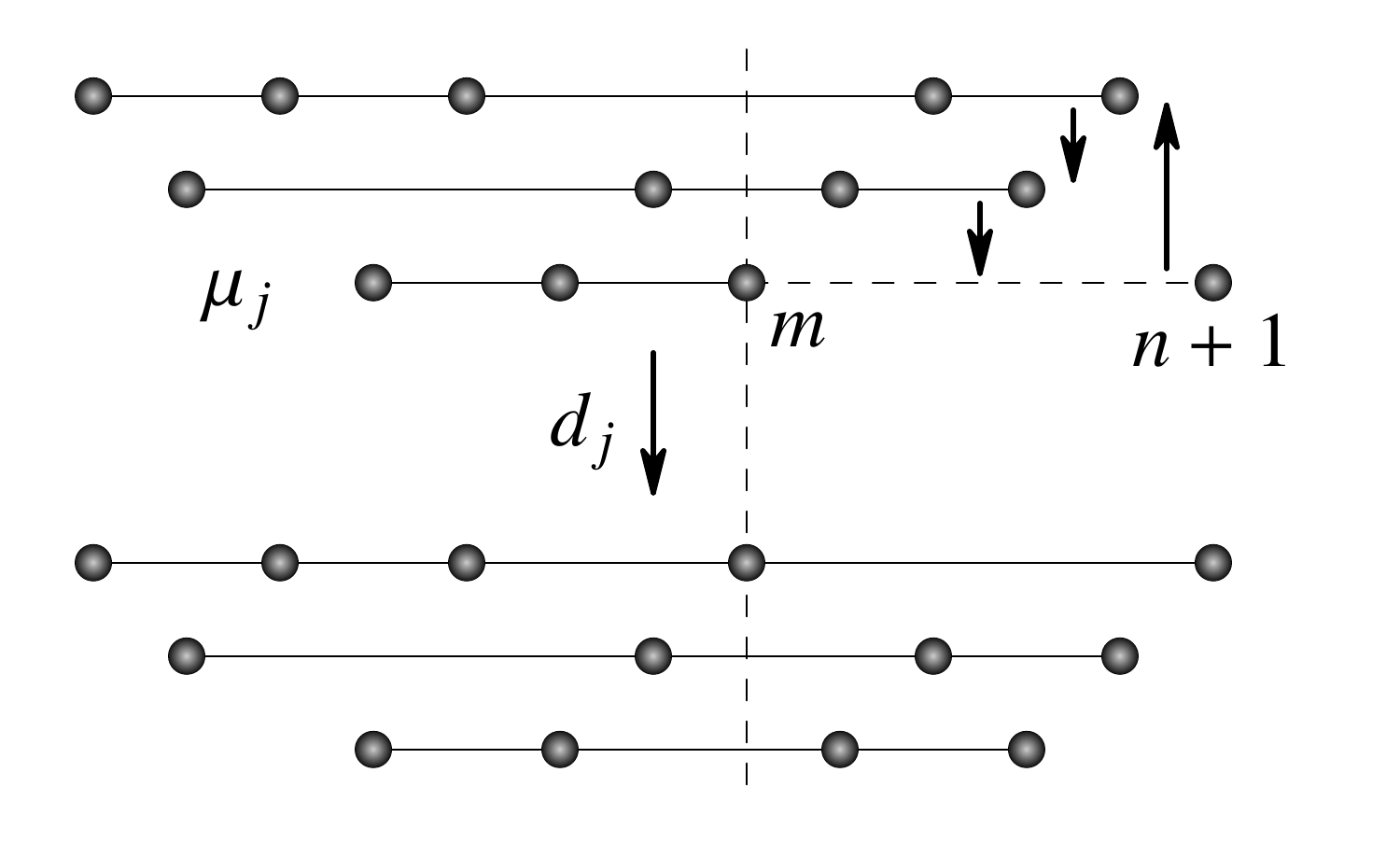}
\caption{Operation $d_j$ for non-overlapping partitions.}
\label{fig:NOP-dj}
\end{figure}

{\em Operation $d_0$}, adding of the singleton $\{n+1\}$ to a partition. This
coincides with the operation $M$ for generic set partitions from section
\ref{s:gf}, but we use another notation and interpret $d_0$ as adding the
element into the team of singletons. Notice, that $d_0\varnothing=\{\{1\}\}$.

{\em Operation $d_j$, $1\le j\le k$}, adding the element $n+1$ to a multiplet.
Let $\mu_1,\dots,\mu_k$ be all multiplets in $\pi\in\Pi^*_{n,k}$, sorted by
increase of their minimal elements. If we just add the element $n+1$ to $\mu_j$
then the multiplets with the supports embracing $\mu_j$ will overlap with this
block (see fig.~\ref{fig:NOP-dj} above). Let these multiplets are those with
the indices $j_1<\dots<j_s=j$. Let us divide them into the left and right parts
with respect to $m=\max\mu_j$:
\[
 \mu^-_{j_r}=\{i\in\mu_{j_r}: i<m\},\quad
 \mu^+_{j_r}=\{i\in\mu_{j_r}: i\ge m\},
\]
and form new multiplets by cyclic permutation of the right parts
(fig.~\ref{fig:NOP-dj} below):
\[
 \tilde\mu_{j_1}=\mu^-_{j_1}\cup\{m,n+1\},\quad
 \tilde\mu_{j_r}=\mu^-_{j_r}\cup\mu^+_{j_{r-1}},~~r=2,\dots,s.
\]
Notice, that the block $\tilde\mu_{j_1}$, which acquires the element $n+1$,
contains 3 elements or more. It is easy to see that the described procedure
brings to the blocks which do not overlap with each other and with the rest
multiplets.

\begin{figure}[bt]
\gr{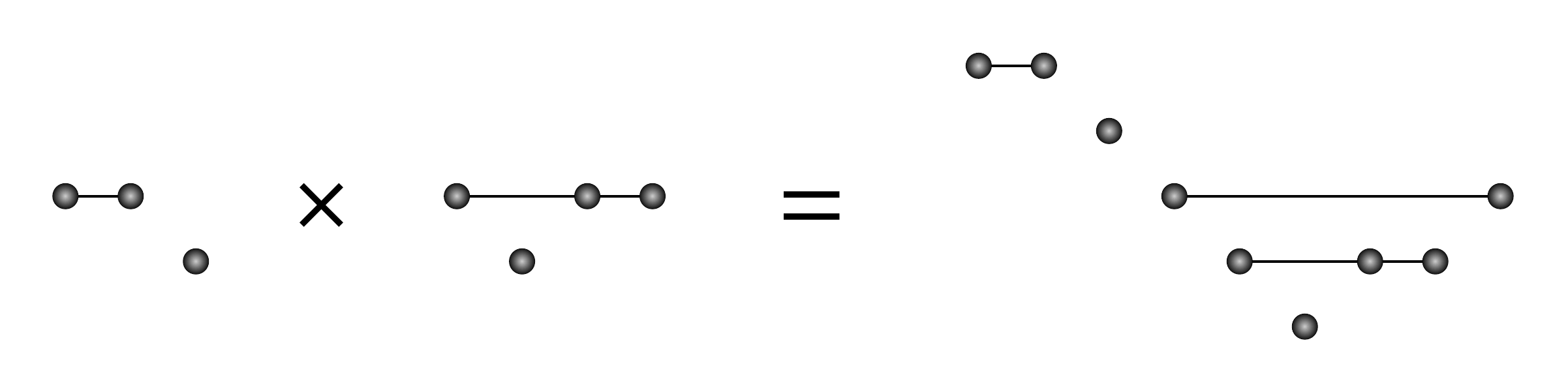}
\caption{Operation $P$ for non-overlapping partitions.}
\label{fig:NOP-P}
\end{figure}

{\em Operation $P$.} Let $\pi\in\Pi^*_{n,k}$, $\rho\in\Pi^*_{m,l}$, then
\begin{equation}\label{NOP.P}
 P(\pi,\rho)=\pi\cup\{\{n+1,m+n+2\}\}\cup(\rho+n+1)\in\Pi^*_{m+n+2,k+l+1}.
\end{equation}
This is a kind of concatenation (\ref{|}), with the wrapping of the second
partition by a doublet (see fig.~\ref{fig:NOP-P}). In particular,
$P(\varnothing,\varnothing)=\{\{1,2\}\}$.

\begin{statement}\label{st:NOP.dP}
Applying all admissible operations $d_j,P$ to the seed partition $\varnothing$
generates all non-overlapping partitions:
\begin{equation}\label{Pi*}
 \Pi^*_0=\{\varnothing\},\quad
 \Pi^*_{n+1}=
 \mathop{\cup}^{\lfloor n/2\rfloor}_{k=0}
   \mathop{\cup}^k_{j=0}d_j\Pi^*_{n,k}
 ~\cup~ \mathop{\cup}^{n-1}_{j=0}P(\Pi^*_j,\Pi^*_{n-1-j}),~~ n\ge0,
\end{equation}
moreover, any partition appears only once.
\end{statement}
\begin{proof}
The last of operations leading to a given partition $\pi\in\Pi^*_{n+1}$ is
defined by the block $\beta$ with the element $n+1$:

--- if $|\beta|=1$ then $\pi=d_0\rho$;

--- if $|\beta|=2$ then $\pi=P(\rho,\sigma)$;

--- if $|\beta|>2$ then $\pi=d_j\rho$, where $j$ is the maximal index of the
multiplet with the support containing the element $m\in\beta$ which precedes
$n+1$.

In all cases, the partitions $\rho$ or $\rho,\sigma$ are determined uniquely
and consist of lesser number of elements, so that the induction can be applied.
\end{proof}

The generating of a given partition is encoded by expression $\varphi(u)$,
builded from a formal variable $u$ by the binary operation $P(a,b)$ and
operations $d_ja$, $0\le j\le\deg a-1$, where $\deg a$ is equal to the number
of occurrences of $u$ in $a$. Let the set $\Phi^*_n$ consist of all such
expressions containing $n+1$ characters $u,d,P$, and the set $\Phi^*_{n,k}$
include those of them which contain $k+1$ characters $u$. Then
\[
 d_j:\Phi^*_{n,k}\to\Phi^*_{n+1,k},~~ 0\le j\le k,\qquad
 P:\Phi^*_{n,k}\times \Phi^*_{m,l}\to\Phi^*_{n+m+2,k+l+1}.
\]
Moreover, $0\le k\le n/2$, because the number of occurrences of $P$ in
$\varphi$ is always one less than the number of occurrences of $u$ (that is, it
is equal to $\deg\varphi-1$). A comparison with the action of the operations on
the partitions leads to the following bijection, taking the Statement
\ref{st:NOP.dP} into account.

\begin{corollary}\label{st:NOP.Phi}
The substitution $\varphi(u)\mapsto\varphi(\varnothing)$ is a one-to-one
mapping $\Phi^*_{n,k}\to\Pi^*_{n,k}$, for all $n,k$.
\end{corollary}

Table \ref{t:Phi*} contains a complete list of expressions $\Phi^*_n$ and
respective partitions $\Pi^*_n$ for $n\le4$. Notice, that $\Pi^*_n=\Pi_n$ at
$n\le3$, and in $\Pi_4$ only the partition $13|24$ is eliminated.

The non-overlapping partitions without singletons $\widetilde\Pi^*_n$ are
builded by the same scheme, but with $d_0$ excluded from the set of generating
operations. The corresponding set $\widetilde\Phi^*_n$ consists of all
expressions $\Phi^*_n$ with no occurrence of $d_0$.

\begin{table}[t]
\[
\begin{array}{|l|l|l|l||l|l|l|l|}
\hline
 n&\quad\varphi(u)&\varphi(\varnothing)&p(\varphi)
&n&\quad\varphi(u)&\varphi(\varnothing)&p(\varphi)\\
\hline
 0& u       & \varnothing & u     &4& d_0d_0d_0d_0u & 1|2|3|4 & u_4   \\
  &               &       &       & & d_0d_0P(u,u)  & 12|3|4  & uu_2  \\
 1& d_0u          & 1     & u_1   & & d_1d_0P(u,u)  & 124|3   & u^2_1 \\
  &               &       &       & & d_0d_1P(u,u)  & 123|4   & u^2_1 \\
 2& d_0d_0u       & 1|2   & u_2   & & d_1d_1P(u,u)  & 1234    & uu_2  \\
  & P(u,u)        & 12    & u^2   & & d_0P(u,d_0u)  & 13|2|4  & u^2_1 \\
  &               &       &       & & d_1P(u,d_0u)  & 134|2   & uu_2  \\
 3& d_0d_0d_0u    & 1|2|3 & u_3   & & d_0P(d_0u,u)  & 1|23|4  & uu_2  \\
  & d_0P(u,u)     & 12|3  & uu_1  & & d_1P(d_0u,u)  & 1|234   & u^2_1 \\
  & d_1P(u,u)     & 123   & uu_1  & & P(u,d_0d_0u)  & 14|2|3  & uu_2  \\
  & P(u,d_0u)     & 13|2  & uu_1  & & P(u,P(u,u))   & 14|23   & u^3   \\
  & P(d_0u,u)     & 1|23  & uu_1  & & P(d_0u,d_0u)  & 1|24|3  & u^2_1 \\
  &               &       &       & & P(d_0d_0u,u)  & 1|2|34  & uu_2  \\
  &               &       &       & & P(P(u,u),u)   & 12|34   & u^3   \\
\hline
\end{array}
\]
\caption{Sets $\Phi^*_n$, $\Pi^*_n$ and monomials $p(\varphi)$ for $n\le4$.}
\label{t:Phi*}
\end{table}

\subsection{Statistics}\label{s:NOP-stat}

In sections \ref{s:SP}, \ref{s:B} we assigned a monomial to any partition by
counting the elements (or positive elements) in the blocks and by multiplying
the corresponding variables $u_l$. For the non-overlapping partitions this does
not bring to a meaningful result, because the operations $d_j$ change the sizes
of blocks unpredictably. In this case, a suitable definition comes from the
encoding of partitions by expressions $\varphi\in\Phi^*_n$. We define the
monomial $p(\varphi)$ as the value of expression $\varphi$, computed according
to the following rules:

(i) operation $d_ja$ adds 1 to the subscript of $u$ for $j+1$-th occurrence of
$u$ in $a$, counting from the left (as usual, we assume that $u=u_0$);

(ii) all operations $P(a,b)$ are applied after $d_j$, and their result is equal
to the product $ab$.

Let us define the polynomials
\begin{equation}\label{fphi}
 f_n(u,u_1,\dots,u_n)=\sum_{\varphi\in\Phi^*_n}p(\varphi)
\end{equation}
by summation over all expressions $\Phi^*_n$, and consider the generating
function
\begin{equation}\label{f}
f=-z/2+f_0z^{-1}+\cdots+f_{n-1}z^{-n}+\cdots\;.
\end{equation}

\begin{statement}\label{st:fn}
Polynomials $f_n$ in variables $u_i=D^i(u)$, $D=\partial_x$, are governed by
recurrence relations
\begin{equation}\label{fn}
 f_0=u,\quad
 f_{n+1}=D(f_n)+\sum^{n-1}_{s=0}f_sf_{n-1-s},\quad n=0,1,2,\dots\;,
\end{equation}
equivalent to the Riccati equation
\begin{equation}\label{Df}
 D(f)+f^2=z^2/4-u.
\end{equation}
\end{statement}
\begin{proof}
Equation (\ref{fn}) follows from (\ref{Pi*}) and identities
\[
 \sum^{\deg\varphi-1}_{j=0}p(d_j\varphi)=D(p(\varphi)),\quad
 p(P(\varphi,\phi))=p(\varphi)p(\phi),
\]
where $\varphi\in\Phi^*_n$, $\phi\in\Phi^*_m$. The second identity follows from
the rule (ii). In order to prove the first one, let us preserve the same order
of variables $u_i$ in a monomial, as they appear after applying of operations
$d_j$ according to the rule (i). Then, if a monomial corresponding to
expression $\varphi\in\Phi^*_{n,k}$ is equal to $p(\varphi)=u_{i_0}\cdots
u_{i_k}$, then $p(d_j\varphi)=u_{i_0}\cdots u_{i_j+1}\cdots u_{i_k}$, $0\le
j\le k$, and we only have to sum over $j$.
\end{proof}

In this proof, the operation $d_j$ is identified with the action of
differentiation $D$ onto one of the factors according to the Leibniz rule. This
leads to the following simple description: {\em in polynomials $f_n$, the
coefficient of any monomial is equal to the number of ways to obtain this
monomial by multiplication and differentiation.} For instance, there are 4 ways
for the monomial $uu_1$ (see table \ref{t:Phi*}): first multiply $u$ by $u$ and
next apply $D$ to the first or second factor, or first apply $D$ to $u$ and
then multiply by $u$ from left or from right. The few first polynomials $f_n$
are written down in table \ref{t:fn}.

Now, let us consider the set $\widetilde\Pi^*_n$ of non-overlapping partitions
without singletons. Since the corresponding expressions $\varphi$ do not
contain $d_0$, hence the first occurrence of $u$ is not differentiated when
computing the monomial $p(\varphi)$, and therefore $p(\varphi)$ is divisible by
$u$. So, let us assign to these partitions the polynomial
\[
 \tilde f_n(u,\dots,u_n)=\frac{1}{u}\sum_{\varphi\in\widetilde\Phi^*_n}p(\varphi).
\]
In particular, one finds easily from table \ref{t:Phi*}
\[
 \tilde f_0=1,\quad \tilde f_1=0,\quad \tilde f_2=u,\quad
 \tilde f_3=u_1,\quad \tilde f_4=u_2+2u^2.
\]
It turns out that these polynomials are very simply related with the
polynomials for all non-overlapping partitions.

\begin{table}[t]
\begin{align*}
 f_0 &= u\\
 f_1 &= u_1\\
 f_2 &= u_2+u^2\\
 f_3 &= u_3+4uu_1\\
 f_4 &= u_4+(6uu_2+5u^2_1)+2u^3\\
 f_5 &= u_5+(8uu_3+18u_1u_2)+16u^2u_1\\
 f_6 &= u_6+(10uu_4+28u_1u_3+19u^2_2)+(30u^2u_2+50uu^2_1)+5u^4
\end{align*}
\[
\begin{array}{c|lllllllll}
n\backslash k
    & 0 & 1    & 2     & 3     & 4    & 5 &~& B^*_n & \widetilde B^*_n\\
\cline{1-7}
 0  & 1 &      &       &       &      &    && 1     & 1   \\
 1  & 1 &      &       &       &      &    && 1     & 0   \\
 2  & 1 & 1    &       &       &      &    && 2     & 1   \\
 3  & 1 & 4    &       &       &      &    && 5     & 1   \\
 4  & 1 & 11   & 2     &       &      &    && 14    & 3   \\
 5  & 1 & 26   & 16    &       &      &    && 43    & 7   \\
 6  & 1 & 57   & 80    & 5     &      &    && 143   & 20  \\
 7  & 1 & 120  & 324   & 64    &      &    && 509   & 60  \\
 8  & 1 & 247  & 1170  & 490   & 14   &    && 1922  & 195 \\
 9  & 1 & 502  & 3948  & 2944  & 256  &    && 7651  & 675 \\
 10 & 1 & 1013 & 12776 & 15403 & 2730 & 42 && 31965 & 2480
\end{array}
\]
\captionsetup{width=0.95\textwidth}
\caption{Polynomials $f_n$, homogeneity $w(u_j)=j+2$. Number triangle
$|\Pi^*_{n,k}|$ for non-overlapping partitions of $[n]$ with $k$ multiplets.
The Bessel numbers $B^*_n=|\Pi^*_n|$ are equal to the sums over the rows;
$\widetilde B^*_n=|\widetilde\Pi^*_n|$.}
\label{t:fn}
\end{table}

\begin{statement}\label{st:tfn}
The generating function for $\tilde f_n$ is equal to
\begin{equation}\label{tf}
 \tilde f=\tilde f_0z^{-1}+\tilde f_1z^{-2}+\cdots+\tilde f_{n-1}z^{-n}+\cdots
 =\frac{1}{z/2-f}.
\end{equation}
\end{statement}
\begin{proof}
By repeating the arguments in the previous proof, we obtain that $\tilde
f=\tilde f_0z^{-1}+\tilde f_1z^{-2}+\cdots$ satisfies the equation $D(\tilde
f)+u\tilde f^2-z\tilde f+1=0$. From here, all $\tilde f_n$ are determined
uniquely through recurrence relations which can be easy written down. The
change $f=z/2-1/\tilde f$ brings to equation (\ref{Df}), and since the leading
term of $f$ is equal to $z/2-1/(1/z)=-z/2$, hence $f$ coincides with the series
(\ref{f}).
\end{proof}

Let us pass to the simplified statistics, as in the previous sections, by
merging the terms of the same degree in $f_n$ under the substitution $u_i=u$.
Since $\deg p(\varphi)$ is one more than the number of multiplets in the
corresponding partition $\varphi(\varnothing)$, hence the coefficient of
$u^{k+1}$ is equal to the cardinality of the set $\Phi^*_{n,k}$, or, what is
the same, of the set $\Pi^*_{n,k}$.

\begin{corollary}\label{st:|Pi*nk|}
The number $|\Pi^*_{n,k}|$ of non-overlapping partitions of $[n]$ with $k$
multiplets is equal to the coefficient of $u^{k+1}$ in the polynomial in
one variable $F_n(u)$, defined by recurrence relations
\begin{equation}\label{Fn}
 F_0=u,\quad
 F_{n+1}=uF'_n+\sum^{n-1}_{s=0}F_sF_{n-1-s},\quad n=0,1,2,\dots\;.
\end{equation}
The generating function $F=-z/2+\sum_{n\ge1}F_{n-1}z^{-n}$ satisfies the
equation
\begin{equation}\label{F'}
 uF'(u)+F(u)^2=z^2/4-u.
\end{equation}
\end{corollary}

\begin{corollary}\label{st:|tPi*nk|}
The number $|\widetilde\Pi^*_{n,k}|$ of non-overlapping partitions $[n]$ with
$k$ multiplets and without singletons is equal to the coefficient of $u^k$ in
the polynomial $\widetilde F_n(u)$, where
\[
 \sum_{n\ge1}\widetilde F_{n-1}z^{-n}=\frac{1}{z-\sum_{n\ge1}F_{n-1}z^{-n}}.
\]
\end{corollary}

The substitution $u=\frac{y^2}{4}$, $F(u)=\frac{y\psi'(y)}{2\psi(y)}$ reduces
(\ref{F'}) to the Bessel equation
\[
 y^2\psi''(y)+y\psi'(y)+(y^2-z^2)\psi(y)=0
\]
with the linearly independent solutions $\psi=J_{\pm z}(y)$, where $J_z$ is the
Bessel function
\[
 J_z(y)=\sum^\infty_{k=0}\frac{(-1)^k(y/2)^{z+2k}}{k!\Gamma(z+k+1)}.
\]
This function satisfies the recurrence relations
\[
 J_{z-1}(y)=J'_z(y)+\frac{z}{y}J_z(y),\quad
 J_{z+1}(y)-\frac{2z}{y}J_z(y)+J_{z-1}(y)=0.
\]
Generating functions $f,F,\tilde f,\widetilde F$ satisfy the corresponding
Riccati equations, as formal asymptotic series with respect to the parameter
$z$. Taking into account that $F(u)=-z/2+0z^0+\dots$, we obtain the asymptotic
expansion
\begin{equation}\label{FJ}
 \frac{yJ'_z(y)}{2J_z(y)}=-\frac{z}{2}+\frac{yJ_{z-1}(y)}{2J_z(y)}
 \sim F\bigl(\frac{y^2}{4},-z\bigr),\quad\mathop{\rm Re}z\to+\infty.
\end{equation}

{\em Bessel numbers} $B^*_n=|\Pi^*_n|=F_n(1)$ counting the non-overlapping
partitions of $[n]$ are equal to the total sums of the coefficients of $f_n$.
Analogously, {\em 2-associated Bessel numbers} $\widetilde
B^*_n=|\widetilde\Pi^*_n|=\widetilde F_n(1)$ count the non-overlapping
partitions without singletons. For the first time, both sequences were defined
in \cite{Flajolet_Schott_1990}, where an asymptotic expansion was obtained
which coincides with (\ref{FJ}) under the substitution $y=2$. However, its
derivation was based on a different combinatorial technique: instead of the
above generating operations which bring to the Riccati equation, the so-called
path diagrams were used \cite{Flajolet_1980} which lead to a representation of
$F$ as a continued fraction, which is equivalent to the three-term recurrence
relation for $J_z$. From here, a functional equation for the generating
function $F(1,z)$ can be derived, which implies a recurrence relation
intermediately for the numbers $B^*_n$ \cite[eq.(22)]{Klazar_2003}, see also
\cite{Claesson_2001}, \citeo{A006789} and \citeo{A099950} (sequence $\widetilde
B^*_n+\widetilde B^*_{n+1}$). This is equivalent to the following statement
after elimiantion of $\widetilde B^*_n$.

\begin{statement}\label{st:BB*}
Sequences $B^*_n,\widetilde B^*_n$ satisfy the recurrence relations
\begin{equation}\label{BB*}
 \widetilde B^*_0=1,\quad
 B^*_n=\sum^n_{j=0}\binom{n}{j}\widetilde B^*_{n-j},\quad
 \widetilde B^*_{n+1}=\sum^{n-1}_{j=0}\widetilde B^*_jB^*_{n-1-j},\quad n\ge0.
\end{equation}
\end{statement}
\begin{proof}
The equation for $\widetilde B^*_{n+1}$ follows from the Corollary
\ref{st:|tPi*nk|}, the equation for $B^*_n$ appears if we enlarge the
partitions without singletons until the necessary size, by inserting singletons
in all possible ways.
\end{proof}

The diagonal entries of the number triangle in table \ref{t:fn}, that is, the
coefficients of ``dispersionless terms'' $u^{k+1}$ in $f_{2k}$, form the
sequence of Catalan numbers \citeo{A000108}. These monomials correspond to
partitions builded only by operation $P$, that is, consisting only from
non-overlapping doublets, which can be easily identified with the strings of
balanced parentheses. Correspondingly, the replacement of the initial condition
in equation (\ref{fn}) by 1 leads to the recurrence for the Catalan numbers
interpolated with zeroes \citeo{A126120}:
\[
 c_0=1,~~ c_{n+1}=\sum^{n-1}_{s=0}c_sc_{n-1-s}
 ~~\to~~ 1,0,1,0,2,0,5,0,~\dots\;.
\]

\subsection{Korteweg--de Vries hierarchy}\label{s:KdV}

Recall that the Riccati equation (\ref{Df}) plays the key role in the theory of
KdV equation
\[
 u_{,t_3}=u_3+6uu_1,
\]
as a tool for computing both conservation laws and higher symmetries
\cite{Gelfand_Dikii_1975}. The KdV hierarchy is defined by the compatibility
conditions for the Schr\"odinger equation
\begin{equation}\label{KdV.psi}
 D^2(\psi)=(z^2/4-u)\psi
\end{equation}
and equations of the form
$\psi_{,t_{2n+1}}=G_{2n}D(\psi)-\frac{1}{2}D(G_{2n})\psi=0$, where
$G_{2n}=z^{2n}+2g_0z^{2n-2}+\cdots+2g_{2n-2}$. A straightforward computation
brings to equations $u_{,t_{2n+1}}=D(g_{2n})$ and the recurrence relations
\begin{equation}\label{gRec}
 g_0=u_0,\quad D(g_{2n+2})=D^3(g_{2n})+4uD(g_{2n})+2u_1g_{2n},\quad n\ge0.
\end{equation}
This implies the equation for the generating function
$g=1+2\sum_{n\ge0}g_{2n}z^{-2n-2}$
\begin{equation}\label{KdV.g}
 2gD^2(g)-D(g)^2+(4u-z^2)g^2=-z^2,
\end{equation}
which uniquely defines all $g_{2n}$ as polynomials, homogeneous with respect to
the weight $w(u_i)=i+2$ (the right hand side of (\ref{KdV.g}) can be replaced
with arbitrary constant series in powers of $z^{-2}$, then the homogeneous
flows are replaced by their linear combinations). It is easy to check that
solution of equation (\ref{KdV.g}) is expressed through solutions of equation
(\ref{Df}):
\begin{equation}\label{gf}
 g=\frac{z}{f(-z)-f(z)}\;,
\end{equation}
which is equivalent to equations
\[
 g_0=f_0=u,\quad
 g_{2n}=f_{2n}+2\sum^{n-1}_{s=0}g_{2s}f_{2n-2-2s},\quad n=1,2,\dots\;,
\]
complementary to recurrence relations (\ref{fn}). Equation (\ref{Df}) is
brought to (\ref{KdV.psi}) by substitution $f=D(\psi)/\psi$ and, therefore,
equation (\ref{gf}) is equivalent to relation $g=\mathop{\rm
const}\psi(z)\psi(-z)$. The equalities
\[
 f_{,t_{2n+1}}=D(G_{2n}f)-\frac{1}{2}D^2(G_{2n}),\quad
 f(z)+f(-z)=-D(\log g)
\]
demonstrate that $f$ is a generating function for the densities of conservation
laws, and the densities $f_{2n+1}$ are trivial.

Although the relation between the generating functions $f$ and $g$ is very
simple and looks like equation (\ref{tf}), no combinatorial interpretation is
known for the polynomials $g_{2n}$ or recurrence relations (\ref{gRec}), in
terms of non-overlapping partitions or any other combinatorial objects.
Nevertheless, there exist explicit, although rather complicated, expressions
for the coefficients of these polynomials. One of them was obtained already in
\cite{Gelfand_Dikii_1975}, it defines the coefficient of a given monomial as a
certain multiple integral. The structure of expressions from
\cite{Schimming_1995, Avramidi_Schimming_2000, Polterovich_1999} is more
combinatorial.

\section{Atomic partitions}\label{s:AP}

\subsection{Generating operations}\label{s:AP-gen}

Atomic partitions were introduced in the study of the algebra of symmetric
polynomials in noncommutative variables (these set partitions enumerate a
certain basis of generators for this algebra) \cite{Bergeron_Zabrocki_2009}.
Like in the case of non-overlapping partitions, the definition makes use of the
ordering of the underlying set.

\begin{definition}\label{def:AP}
A partition $\pi=\{\pi_1,\dots,\pi_k\}\vdash[n]$ is called {\em atomic}, if no
subset of its blocks forms a partition of $[m]$ for $1\le m<n$.
\end{definition}

Equivalently, no non-empty partitions $\rho,\sigma$ exist such that
$\pi=\rho|\sigma$, where $|$ is the concatenation (\ref{|}). Yet another
equivalent definition reads: the set partition is atomic if the supports of its
blocks cover the whole interval of the partition,
\begin{equation}\label{AP.supp}
 \mathop{\cup}_{j=1}^k \supp\pi_j=[1,n]_\Real.
\end{equation}
Let $\Pi^a_n$ and $\Pi^a_{n,k}$ denote the sets of all atomic partitions of
$[n]$ and those which consist of $k$ blocks, $k=1$ at $n=1$ and $1\le k\le n-1$
at $n>1$.

\begin{figure}[t]
\gr{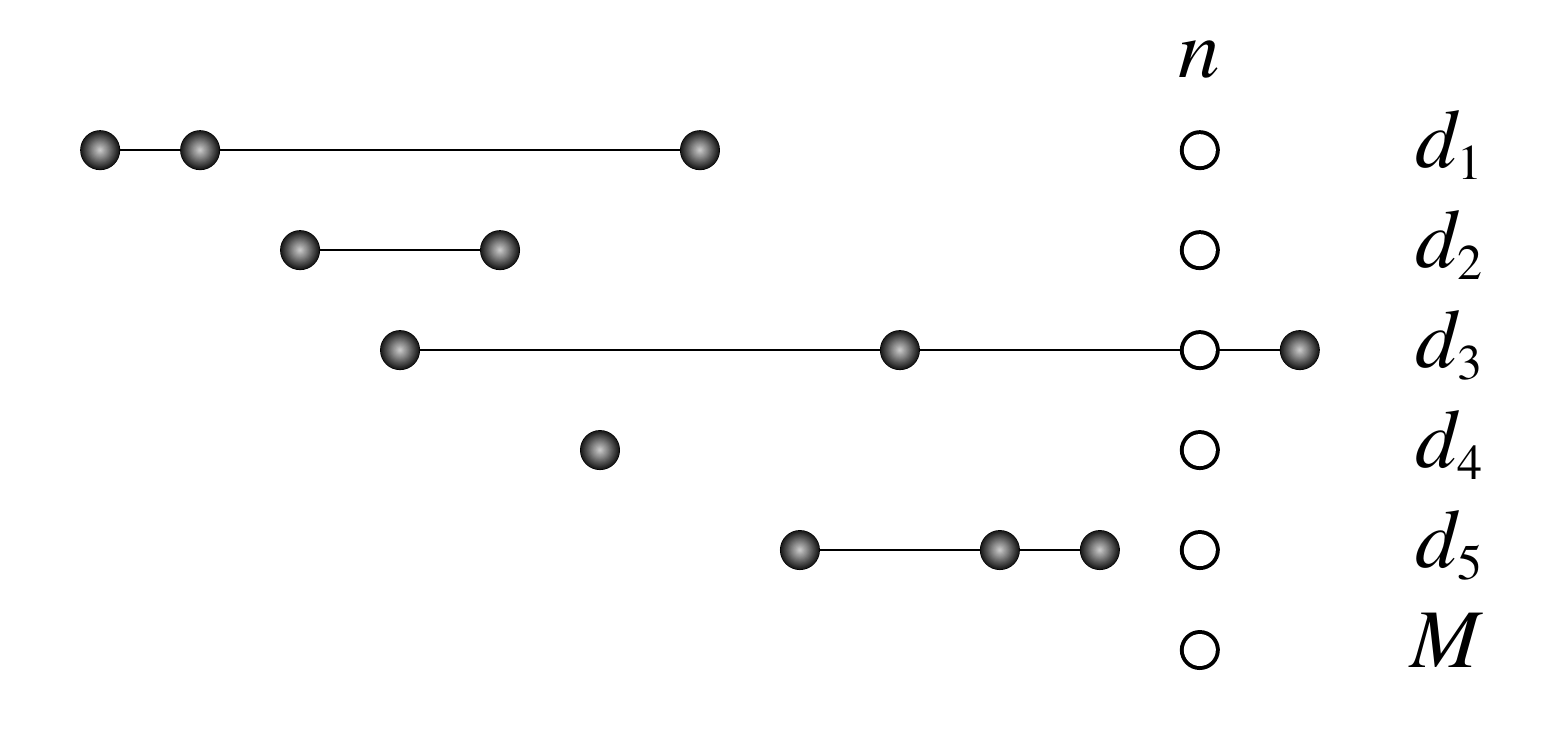}
\caption{Operations $d_j$, $M$ for atomic partitions. Empty circles mark the
vacancies for new element.}
\label{fig:AP-djM}
\end{figure}

\begin{figure}[t]
\gr{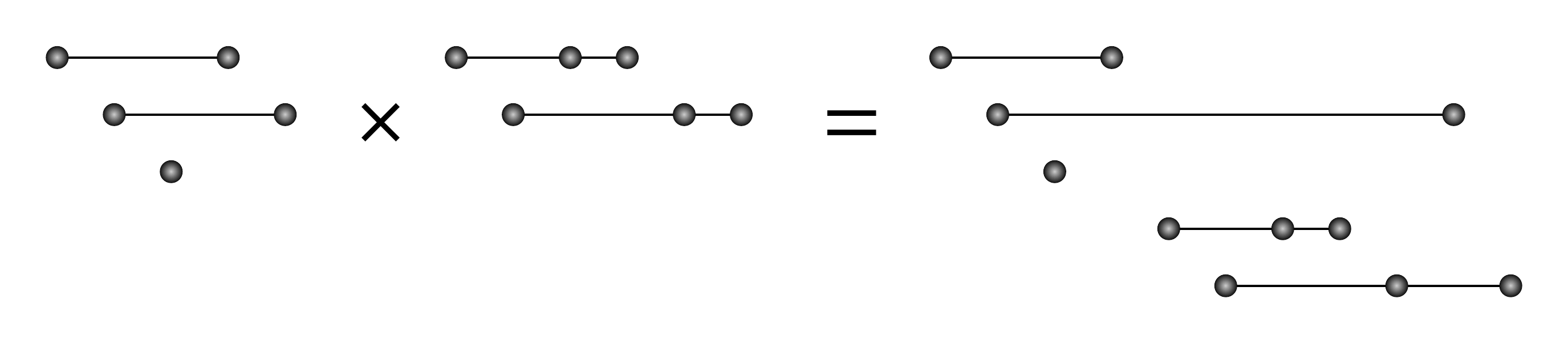}
\caption{Operation $P$ for atomic partitions.}
\label{fig:AP-P}
\end{figure}

Construction of atomic partitions employs the operations
\begin{gather*}
 d_j: \Pi^a_{n,k}\to \Pi^a_{n+1,k},\quad 1\le j\le k,\qquad
 M: \Pi^a_{n,k}\to \Pi^a_{n+1,k+1},\quad n\ge2,\\
 P: \Pi^a_{n,k}\times \Pi^a_{m,l}\to \Pi^a_{m+n,k+l},\quad m,n\ge2.
\end{gather*}
Two first operations are similar to (\ref{SP.djM}), but the difference is that
the new element is added at the $n$-th position, displacing the old element
from this position to $n+1$-st position, cf. fig.~\ref{fig:P-djM} and
\ref{fig:AP-djM}; moreover, operation $M$ is applied only for $n>1$. In more
details, we first replace the element $n$ by $n+1$ in the partition
$\pi=\{\pi_1,\dots,\pi_k\}\vdash[n]$:
\[
 \pi'=\{\pi'_1,\dots,\pi'_k\}=\pi|_{n\to n+1},
\]
next, operation $d_j$ adds the element $n$ to the block $\pi'_j$ (in
particular, if $\pi_j$ contains $n$ then this operation just adds $n+1$ to this
block); operation $M$ adds $n$ as a new singleton:
\begin{equation}\label{AP.djM}
 d_j\pi=\{\pi'_1,\dots,\pi'_j\cup\{n\},\dots,\pi'_k\};\quad
 M\pi=\{\pi'_1,\dots,\pi'_k,\{n\}\},~~ n\ge2.
\end{equation}
Operation $P$ is a modification of concatenation, with displacing of the last
elements of both partitions (cf. with (\ref{NOP.P})). Let $\rho\in\Pi^a_n$,
$\sigma\in\Pi^a_m$, $m,n\ge2$, then
\begin{equation}\label{AP.P}
 P(\rho,\sigma)=\rho|_{n\to m+n-1}\cup(\sigma|_{m\to m+1}+n-1).
\end{equation}
In other words, we first concatenate the partitions with the maximal elements
removed, then we place these elements back to their blocks, in the last but one
position for the block from the first partition and in the last position for
the block from the second one, see fig.~\ref{fig:AP-P}.

\begin{statement}\label{st:AP.djMP}
Operations (\ref{AP.djM}), (\ref{AP.P}) generate all atomic partitions by
application to the seed partition $\{\{1\}\}$ in all possible ways:
\begin{gather}
\nonumber
 \Pi^a_1=\{\{\{1\}\}\},\quad \Pi^a_2=d_1\Pi^a_1=\{\{\{1,2\}\}\},\\
\label{Pi^a}
 \Pi^a_{n+1}=\mathop{\cup}^{n-1}_{k=1}\mathop{\cup}^k_{j=1}d_j\Pi^a_{n,k}
 ~\cup~ M\Pi^a_n ~\cup~
 \mathop{\cup}^{n-1}_{j=2}P(\Pi^a_j,\Pi^a_{n+1-j}),~~ n\ge2,
\end{gather}
moreover, any partition appear only once.
\end{statement}
\begin{proof}
Let us prove that our operations preserve the atomicity property. Notice, that
in partition $\pi\in\Pi^a_n$, the block $\pi_j$ containing $n$ is not a
singleton if $n>1$, since otherwise the partition is of the form
$\pi=\rho|\{\{1\}\}$. Therefore, the passage to the partition $\pi'=\pi|_{n\to
n+1}$ enlarges the support of this block:
$\supp\pi'_j=\supp\pi_j\cup[n,n+1]_\Real$. The further adding of the element
$n$ does not reduce the blocks supports and it follows by (\ref{AP.supp}) that
partitions $d_j\pi$, $M\pi$ are atomic as well. The exceptional case is $n=1$,
for which $M\{\{1\}\}=\{\{1\},\{2\}\}$, this is why the definition of this
operation includes the restriction $n\ge2$.

In (\ref{AP.P}), analogously, the blocks of partitions $\rho|_{m\to m+n-1}$ and
$\sigma|_{n\to n+1}+m-1$ cover, respectively, the intervals $[1,m+n-1]_\Real$
and $[m,m+n]_\Real$ (again, take into account that if $m,n\ge2$ then the blocks
with maximal elements are not singletons), and their union covers
$[1,m+n]_\Real$.

Now let us prove that any partition $\pi\in\Pi^a_{n+1}$ can be represented, in
a unique way, as $\pi=M\tau$ or $\pi=d_j\tau$ or $\pi=P(\rho,\sigma)$, where
$\rho,\sigma,\tau$ are atomic partitions. Let $n$ (last to the end element of
$\pi$) belongs to the block $\pi_j$. Construct the partition $\tau\vdash[n]$ by
deleting this element and moving $n+1$ at the vacant position. The following
cases are possible.

1) $\pi_j$ is a singleton. Then $\tau\in\Pi^a_n$ and $\pi=M\tau$; moreover,
$\pi$ cannot be obtained in other way, because the operations $d_j$, $P$ add
the element $n$ to a block which already contains some elements.

2) $\pi_j$ is not a singleton and $\tau\in\Pi^a_n$. Then $\pi=d_j\tau$; the
uniqueness follows from the fact that if $\pi=P(\rho,\sigma)$ then
$\tau\not\in\Pi^a_n$ and $j$ is uniquely determined as the index of a block
containing the element $n$.

3) $\pi_j$ is not a singleton and $\tau\not\in\Pi^a_n$. Then $\tau$ is uniquely
represented as $\tau=\tilde\rho|\sigma$, where $\sigma\in\Pi^a_s$, $s>1$. The
partition $\sigma$ is characterized by the property that the union of the
supports of $\sigma+|\tilde\rho|$ covers the interval to the right of the
rightmost gap in the union of the supports of $\tau$. Moreover, the block
$\pi_j\setminus\{n\}$ belongs to the partition $\tilde\rho$ and appending of
$|\tilde\rho|+1$ to this block gives us an atomic partition $\rho$ such that
$\pi=P(\rho,\sigma)$.
\end{proof}

\subsection{Statistics}\label{s:AP-stat}

\paragraph{\mdseries\em 1) Back to the Bell polynomials.} We will consider two
statistics for the atomic partitions. First, let us copy the definition of the
Bell polynomials $Y_n$ (\ref{Yn}):
\begin{equation}\label{tYn}
 p(\pi)=\prod^{|\pi|}_{j=1}v_{|\pi_j|},\qquad
 \widetilde Y_n(v_1,\dots,v_n)=\sum_{\pi\in\Pi^a_n}p(\pi).
\end{equation}
It turns out that polynomials $\widetilde Y_n$ and $Y_n$ are closely related
(cf. with Statement \ref{st:tfn}). Pay attention that equation (\ref{tYY})
involves the ordinary generating functions, in contrast to the exponential one
in (\ref{Yn-egf}). The differential equations for this functions can be solved
in quadratures, but these formulae are of little value for us, since $Y$ and
$\widetilde Y$ serve just as asymptotic expansions for solutions, like the
series in section \ref{s:NOP-stat}.

\begin{statement}\label{st:tYY}
Generating functions $\widetilde Y=\widetilde Y_1/z+\widetilde Y_2/z^2+\cdots$
and $Y=Y_0+Y_1/z+Y_2/z^2+\cdots$ are related by equation
\begin{equation}\label{tYY}
 \widetilde Y=1-1/Y.
\end{equation}
\end{statement}
\begin{proof}
Let $v_i=D^i(v)$, $D=\partial_x$. It is easy to prove that the generating
operations (\ref{AP.djM}), (\ref{AP.P}) satisfy the property
\[
 \sum^{|\pi|}_{j=1}p(d_j\pi)=D(p(\pi)),\quad p(M\pi)=v_1p(\pi),\quad
 p(P(\rho,\sigma))=p(\rho)p(\sigma).
\]
Statement \ref{st:AP.djMP} implies the recurrence relations
\begin{equation}\label{tYn-rec}
 \widetilde Y_1=v_1,\quad \widetilde Y_2=v_2,\quad
 \widetilde Y_{n+1}=D(\widetilde Y_n)+v_1\widetilde Y_n
 +\sum^{n-1}_{s=2}\widetilde Y_s\widetilde Y_{n+1-s},~ n\ge2,
\end{equation}
which are equivalent to the Riccati equation
\begin{equation}\label{DtY}
 D(\widetilde Y)+(\widetilde Y-1)(z\widetilde Y-v_1)=0.
\end{equation}
The change (\ref{tYY}) brings to equation $D(Y)=(z-v_1)Y-z$ which is equivalent
to the recurrence relations (\ref{Yn-rec}) for the Bell polynomials.
\end{proof}

Let $a_{n,k}=|\Pi^a_{n,k}|$ denote the number of atomic partitions $[n]$ with
$k$ blocks \citeo{A087903} and $a_n=|\Pi^a_n|$ be the total number of atomic
partitions \citeo{A074664}, see table \ref{t:rn}. The substitution of $v_j=v$
and $v_j=1$ into (\ref{tYY}) establishes the connection of these numbers with
the Stirling numbers of the 2nd kind and the Bell numbers:
\[
 \sum_{n\ge1}\sum^n_{k=1}a_{n,k}\frac{v^k}{z^n}
  =1-\biggl(\sum_{n\ge0}\sum^n_{k=0}
    \StirlingII{n}{k}\frac{v^k}{z^n}\biggr)^{-1},\quad
 \sum_{n\ge1}\frac{a_n}{z^n}=1-\biggl(\sum_{n\ge0}\frac{B_n}{z^n}\biggr)^{-1}.
\]
For the first time, these relations were found in
\cite[eq.(1),(2)]{Bergeron_Zabrocki_2009}.

\begin{table}[t]
\begin{align*}
 r_2 &= u\\
 r_3 &= u_1+uv\\
 r_4 &= u_2+(2u_1v+uv_1+u^2)+uv^2\\
 r_5 &= u_3+(3u_2v+4uu_1+3u_1v_1+uv_2)+(3u_1v^2+3uvv_1+3u^2v)+uv^3\\
 r_6 &= u_4+(4u_3v+6uu_2+6u_2v_1+5u^2_1+4u_1v_2+uv_3)\\
     &\qquad +(6u_2v^2+16uu_1v+12u_1vv_1+2u^3+5u^2v_1+4uvv_2+3uv^2_1)\\
     &\qquad +(4u_1v^3+6u^2v^2+6uv^2v_1)+uv^4
\end{align*}
\[
\begin{array}{c|llllllllll}
n\backslash k
   & 1 & 2    & 3     & 4     & 5    & 6  & 7  & 8&~& a_n\\
\cline{1-9}
 1 & 1 &      &       &       &      &    &    &   && 1   \\
 2 & 1 &      &       &       &      &    &    &   && 1   \\
 3 & 1 & 1    &       &       &      &    &    &   && 2   \\
 4 & 1 & 4    & 1     &       &      &    &    &   && 6   \\
 5 & 1 & 11   & 9     & 1     &      &    &    &   && 22  \\
 6 & 1 & 26   & 48    & 16    & 1    &    &    &   && 92  \\
 7 & 1 & 57   & 202   & 140   & 25   & 1  &    &   && 426 \\
 8 & 1 & 120  & 747   & 916   & 325  & 36 & 1  &   && 2146\\
 9 & 1 & 247  & 2559  & 5071  & 3045 & 651& 49 & 1 && 11624
\end{array}
\]
\captionsetup{width=0.95\textwidth}
\caption{Polynomials $r_n$. The numbers $a_{n,k}$ of atomic partitions $[n]$
with $k$ blocks; sums over rows give the total numbers of atomic partitions
$a_n$.}\label{t:rn}
\end{table}

\begin{remark}
The so-called unsplittable partitions with a close statistics were studied in
\cite{Bergeron_Reutenauer_Rosas_Zabrocki_2008, Can_Sagan_2011}. The total
number of such partitions of size $n$ coincides with the number of atomic
partitions, but another number triangle appears instead of $a_{n,k}$. A rather
complicated bijection between these two classes of partitions was established
in \cite{Chen_Li_Wang_2011}. It is not clear at the moment, whether this
correspondence is described by some transformation of generating functions or
variables $v_i$.
\end{remark}

\paragraph{\mdseries\em 2) Polynomials in two variables.}

Let us consider expressions $\varphi(u)$, builded from the variable $u$ by use
of operations $P(a,b)$, $Ma$ and $d_ja$, $1\le j\le\deg a$, where $\deg a$ is
equal to the common number of occurrences of $u$ and $M$ in $a$. Let the set
$\Phi^a_n$, $n\ge2$, consists of all such expressions containing $n-1$
characters $u,d,M,P$, and the set $\Phi^a_{n,k}$ includes those with the degree
equal to $k$. According to Statement \ref{st:AP.djMP}, the map
$\varphi(u)\mapsto\varphi(\{\{1,2\}\})$ defines a bijection between
$\Phi^a_{n,k}$ and $\Pi^a_{n,k}$, for $n\ge2$ (notice, that the partition
$\{\{1\}\}$ which makes up $\Pi^a_1$, is actually not involved into
(\ref{Pi^a})).

In addition to the variable $u=u_0$, we introduce the variable $v=v_0$ and
define the monomial $q(\varphi)$, as the value of expression $\varphi$ computed
according to the following rules, applied in the given order:

(i) all operations $Ma$ are replaced with $P(a,v)$;

(ii) the operation $d_ja$ adds 1 to the subscript of the $j$-th occurrence of
$u$ or $v$ into $a$, counting from the left;

(iii) all operations $P(a,b)$ are replaced with $ab$.

Summation over all expressions defines the polynomials
\begin{equation}\label{rphi}
 r_n(u,\dots,u_{n-2},v,\dots,v_{n-3})=\sum_{\varphi\in\Phi^a_n}q(\varphi),
\end{equation}
see table \ref{t:Phi^a}, where the list of expressions $\Phi^a_n$ is given for
$n=2,3,4$, along with the respective partitions and monomials, and table
\ref{t:rn} containing few first polynomials $r_n$.

\begin{table}[t]
\[
\begin{array}{|l|l|l|l|}
\hline
 n&\varphi(u)   &\varphi(\{\{1,2\}\})&q(\varphi)\\
\hline
 2& u           & 12     & u    \\[3pt]
 3& d_1u        & 123    & u_1  \\
  & Mu          & 13|2   & uv   \\[3pt]
 4& d_1d_1u     & 1234   & u_2  \\
  & d_1Mu       & 134|2  & u_1v \\
  & d_2Mu       & 14|23  & uv_1 \\
  & Md_1u       & 124|3  & u_1v \\
  & MMu         & 14|2|3 & uv^2 \\
  & P(u,u)      & 13|24  & u^2\\
\hline
\end{array}
\]
\caption{The sets $\Phi^a_n$, $\Pi^a_n$ and monomials $q(\varphi)$ for $n=2,3,4$.}
\label{t:Phi^a}
\end{table}

\begin{remark}
A comparison of tables \ref{t:Phi*} and \ref{t:Phi^a} demonstrates that
non-overlapping partitions $\Pi^*_n$ are in one-to-one correspondence with the
atomic partitions $\Pi^a_{n+2}$ constructed without the use of operation $M$.
\end{remark}

Notice, that in partition $P(\rho,\sigma)$ the blocks of $\rho$ and $\sigma$
are enumerated sequentially, and operation $M$ adds singleton as the last
block. On the step (i), the variable $v$ is introduced as the second argument,
hence after this step the enumeration of variables $u,v$ from left to right is
consistent with the enumeration of blocks in respective partition.

Therefore, after the step (ii), a block with $l$ elements corresponds to
variable $u_{l-2}$ or $v_{l-1}$, depending on whether this block was generated
by operation $P$ or $M$. This implies the relation between the monomials $p$
and $q$, and therefore, between polynomials $\widetilde Y_n$ and $r_n$:
\[
 p(\varphi(\{\{1,2\}\}))=q(\varphi)|_{u_l=v_{l+2},v_l=v_{l+1}},\quad
 \widetilde Y_n=r_n|_{u_l=v_{l+2},v_l=v_{l+1}}
\]
(except for $\widetilde Y_1=v_1$, because the polynomial $r_1$ is not defined).

\begin{statement}\label{st:rn}
The polynomials $r_n$ of the variables $u_i=D^i(u)$, $v_i=D^i(v)$,
$D=\partial_x$, satisfy the recurrence relations
\begin{equation}\label{rn}
 r_2=u,\quad
 r_{n+1}=D(r_n)+vr_n+\sum^{n-1}_{s=2}r_sr_{n+1-s},\quad n\ge2.
\end{equation}
The generating function $r=r_2/z+r_3/z^2+\cdots$ satisfies the relation
\begin{equation}\label{Dr}
 D(r)+r^2+vr+u=zr.
\end{equation}
\end{statement}
\begin{proof}
Equation (\ref{rn}) follows from (\ref{Pi^a}) and identities
\[
 \sum^{\deg\varphi}_{j=1}q(d_j\varphi)=D(q(\varphi)),\quad
 q(M\varphi)=vq(\varphi),\quad q(P(\varphi,\phi))=q(\varphi)q(\phi).
\]
The two last identities are obvious. In order to prove the first, let us write
the variables $u_i$, $v_i$ in the monomial in the same order as they stay after
applying the operations $M$ and $d_j$ according to the rules (i), (ii). Then
the difference between the monomials $q(d_j\varphi)$ and $q(\varphi)$ is that
the subscript of $j$-th factor is increased by 1, and we only have to take the
sum over $j$.
\end{proof}

\subsection{Kaup--Broer hierarchy}\label{s:KB}

Linearization of the Riccati equation (\ref{Dr}) brings to equation
\begin{equation}\label{KB.psi}
 D^2(\psi)+(v-z)D(\psi)+u\psi=0.
\end{equation}
The associated hierarchy of nonlinear equations appears from the compatibility
conditions of (\ref{KB.psi}) with equation $\psi_{,t}=GD(\psi)+H\psi=0$, where
$G$ and $H$ are polynomials in $z$. Straightforward computations brings to
equations
\[
 G=(z^{n-1}g)_{\ge0},\quad H=\frac{1}{2}(z^{n-1}((v-z)g-D(g)))_{\ge0},
\]
where the subscript $_{\ge0}$ denotes the deleting of negative powers of $z$,
and to equation for the generating function $g=1+g_1/z+g_2/z^2+\cdots$
\begin{equation}\label{KB.g}
 2gD^2(g)-D(g)^2+(4u-2v_1-(v-z)^2)g^2=-z^2,
\end{equation}
which uniquely determines all $g_j$. Moreover, the compatibility conditions are
equivalent to the system
\[
 u_{,t_n}=\frac{1}{2}D(D(g_n)-vg_n+g_{n+1}),\quad v_{,t_n}(v)=D(g_n).
\]
The choice of the right hand side in (\ref{KB.g}) provides the homogeneity with
respect to the weight $w(u_i)=i+2$, $w(v_i)=i+1$. The basic equations of the
hierarchy are of the form
\[
 u_{,t_2}=D(u_1+2uv),\quad v_{,t_2}=D(-v_1+v^2+2u).
\]
This is the Kaup--Broer system \cite{Kupershmidt_1985}, one of gauge equivalent
forms of the nonlinear Schr\"odinger equation (the chain of substitution can be
found in \cite[eq. (A)]{Mikhailov_Shabat_Yamilov_1988}). Let us write down also
the 3-rd order symmetry:
\[
 u_{,t_3}=D(u_2+3u^2+3u_1v+3uv^2),\quad v_{,t_3}=D(v_2-3vv_1+v^3+6uv).
\]
A comparison of the Riccati equations (\ref{Dr}) and (\ref{DtY}) proves that
all flows admit the reductions $u=0$ and $u=v_1$, both leading to the Burgers
hierarchy; a comparison with the Riccati equation (\ref{Df}) proves that all
odd flows admit the reduction $v=0$ to the KdV hierarchy.

The generating functions $g$ and $r=r(u,v,z)$ from Statement \ref{st:rn} are
related by the formula $g=z/(\bar r-r)$, where the series
\[
 \bar r=-\frac{u}{r(u,-u_1/u-v,-z)}=z-v+\frac{v_1-u}{z}+\cdots
\]
is another solution of the Riccati equation (\ref{Dr}). Like in the case of
equation (\ref{KdV.g}), an intermediate combinatorial interpretation of
polynomials $g_j$ is unknown.

\subsection*{Acknowledgements}

This work was supported by the Russian Foundation for Basic Researches under
grant 16-01-00289a and the Leading Scientific Schools Program under grant
3139.2014.2.

\addcontentsline{toc}{section}{References}

\end{document}